\documentclass[prx,reprint,a4paper,superscriptaddress,citeautoscript,floatfix,svgnames,nofootinbib]{revtex4-2}

\usepackage{amsmath,amsfonts,amssymb}
\usepackage{newtxtext,newtxmath}
\usepackage{mathrsfs}
\usepackage[pdftex]{graphicx}
\usepackage{microtype}
\usepackage{bm}\let\vec\bm

\usepackage[svgnames]{xcolor}
\usepackage[
	colorlinks=True,linkcolor=DarkRed,citecolor=ForestGreen,urlcolor=MediumBlue,
	pdfstartview=FitH,bookmarks=False,pdfpagemode=UseNone
]{hyperref}

\newcommand{\kB}{k_{\mathrm{B}}}
\def\EDF{Supplementary Fig.}
\def\SIS{Supplementary Information Sec.}
\let\oldaddcontentsline\addcontentsline%
\renewcommand{\addcontentsline}[3]{}%

\begin{document}

\title{Reconciling scaling of the optical conductivity of cuprate superconductors\\with Planckian resistivity and specific heat}
\homepage[This version of the article has been accepted for publication after peer review but is not the Version of Record and does not reflect post-acceptance improvements, or any corrections. The Version of Record is available online at: ]{http://doi.org/10.1038/s41467-023-38762-5}

\author{B. Michon}
\affiliation{Department of Quantum Matter Physics, University of Geneva, 24 quai Ernest-Ansermet, 1211 Geneva, Switzerland}
\affiliation{Department of Physics, City University of Hong Kong, 83 Tat Chee Avenue, Kowloon, Hong Kong, China}
\affiliation{Hong Kong Institute for Advanced Study, City University of Hong Kong, 83 Tat Chee Avenue, Kowloon, Hong Kong, China}
\author{C. Berthod}
\affiliation{Department of Quantum Matter Physics, University of Geneva, 24 quai Ernest-Ansermet, 1211 Geneva, Switzerland}
\author{C. W. Rischau}
\affiliation{Department of Quantum Matter Physics, University of Geneva, 24 quai Ernest-Ansermet, 1211 Geneva, Switzerland}
\author{A. Ataei}
\affiliation{Institut Quantique, D\'{e}partement de Physique \& RQMP, Universit\'{e} de Sherbrooke, Sherbrooke, Qu\'{e}bec, Canada}
\author{L. Chen}
\affiliation{Institut Quantique, D\'{e}partement de Physique \& RQMP, Universit\'{e} de Sherbrooke, Sherbrooke, Qu\'{e}bec, Canada}
\author{S.~Komiya}
\affiliation{Energy Transformation Research Laboratory, Central Research Institute of Electric Power Industry, 2-6-1 Nagasaka, Yokosuka, Kanagawa, Japan}
\author{S. Ono}
\affiliation{Energy Transformation Research Laboratory, Central Research Institute of Electric Power Industry, 2-6-1 Nagasaka, Yokosuka, Kanagawa, Japan}
\author{L. Taillefer}
\affiliation{Institut Quantique, D\'{e}partement de Physique \& RQMP, Universit\'{e} de Sherbrooke, Sherbrooke, Qu\'{e}bec, Canada}
\affiliation{Canadian Institute for Advanced Research, Toronto, Ontario, Canada}
\author{D. van der Marel}
\email{dirk.vandermarel@unige.ch}
\affiliation{Department of Quantum Matter Physics, University of Geneva, 24 quai Ernest-Ansermet, 1211 Geneva, Switzerland}
\author{A. Georges}
\email{antoine.georges@college-de-france.fr}
\affiliation{Coll\`{e}ge de France, 11 place Marcelin Berthelot, 75005 Paris, France}
\affiliation{Center for Computational Quantum Physics, Flatiron Institute, New York, New York 10010, USA}
\affiliation{Department of Quantum Matter Physics, University of Geneva, 24 quai Ernest-Ansermet, 1211 Geneva, Switzerland}
\affiliation{CPHT, CNRS, \'{E}cole Polytechnique, IP Paris, F-91128 Palaiseau, France}

\date{May 26, 2023}

\begin{abstract}
Materials tuned to a quantum critical point display universal scaling properties as a function of temperature $T$ and frequency $\omega$. A long-standing puzzle regarding cuprate superconductors has been the observed power-law dependence of optical conductivity with an exponent smaller than one, in contrast to $T$-linear dependence of the resistivity and $\omega$-linear dependence of the optical scattering rate. Here, we present and analyze resistivity and optical conductivity of La$_{2-x}$Sr$_x$CuO$_4$ with $x=0.24$. We demonstrate $\hbar\omega/\kB T$ scaling of the optical data over a wide range of frequency and temperature, $T$-linear resistivity, and optical effective mass proportional to $\sim \ln T$ corroborating previous specific heat experiments. We show that a $T,\omega$-linear scaling \textit{Ansatz} for the inelastic scattering rate leads to a unified theoretical description of the experimental data, including the power-law of the optical conductivity. This theoretical framework provides new opportunities for describing the unique properties of quantum critical matter.
\end{abstract}

\maketitle

\section{Introduction}

The linear-in-temperature electrical resistivity is one of the remarkable properties of the cuprate high temperature superconductors \cite{Hussey-2008, Proust-2019, Varma-2020, Varma-2002}. By means of chemical doping, it is possible to tune these materials to a carrier concentration where $\rho(T)=\rho_0+AT$ in a broad temperature range. For Bi$_{2+x}$Sr$_{2-y}$CuO$_{6\pm\delta}$, it has been possible to demonstrate this from 7 to 700~K \cite{Martin-1990} by virtue of the low $T_c$ of this material. For the underdoped cuprates, the linear-in-$T$ resistivity is ubiquitous for temperatures $T>T^*$, where $T^*$ is a doping-dependent cross-over temperature that decreases as a function of doping and vanishes at a critical doping $p^*$. From one cuprate family to another, the exact value of $p^*$ varies widely within the range $0.19<p^*<0.40$ \cite{Badoux-2016, Laliberte-2016, Collignon-2017, Putzke-2021, Lizaire-2021}. For doping levels $p<p^*$, many of the physical properties indicate the presence of a pseudogap that vanishes at $p^*$ \cite{Matt-2015, Cyr-Choiniere-2018}. When $p$ is tuned exactly to $p^*$, the $T$-linear resistivity persists down to $T=0$~K if superconductivity is suppressed e.g.\ by applying a magnetic field \cite{Daou-2009, Cooper-2009, Michon-2018}. The conundrum of the $T$-linear resistivity has been associated to the idea that the momentum relaxation rate cannot exceed the Planckian dissipation $\kB T/\hbar$ \cite{Zaanen-2004, Zaanen-2019, Bruin-2013, Hartnoll-2022}, a state of affairs for which there exists now strong experimental support \cite{Legros-2019, Grissonnanche-2021}.

As expected for a system tuned to a quantum critical point \cite{Sachdev-2011}, $\hbar\omega/\kB T$ scaling has been observed in the optical properties of high-$T_c$ cuprates \cite{vanderMarel-2003, vanderMarel-2006} over some range of doping. The optical scattering rate obtained from an extended Drude fit to the data was found to obey a $T$-linear dependence $1/\tau\sim \kB T/\hbar$ in the low-frequency regime ($\hbar\omega\lesssim 1.5 \kB T$) as well as a linear dependence on energy over an extended frequency range \cite{Schlesinger-1990a, Schlesinger-1990b, Cooper-1993, ElAzrak-1994, Baraduc-1996, vanderMarel-2006}. A direct measurement of the linear temperature dependence of the single-particle relaxation rate extending over 70\% of the Fermi surface was obtained with angle resolved photoemission spectroscopy (ARPES) \cite{Valla-2000}. These observations are qualitatively consistent with the $T$-linear dependence of the resistivity and Planckian behavior. In contrast, by analyzing the modulus and phase of the optical conductivity itself, a power-law behavior $\sigma(\omega)=C/(-i\omega)^{\nu^*}$ with an exponent $\nu^*<1$ was reported at higher frequencies $\hbar\omega \gtrsim 1.5 \kB T$ \cite{ElAzrak-1994, Baraduc-1996, Ioffe-1998, vanderMarel-2003, vanderMarel-2006, Hwang-2007}. The exponent was found to be in the range $\nu^*\approx 0.65$ with some dependence on sample and doping level \cite{Schlesinger-1990b, ElAzrak-1994, Baraduc-1996, vanderMarel-2003}. Hence, from these previous analyses, it would appear that different power laws are needed to describe optical spectroscopy data: one at low frequency consistent with $\hbar\omega/\kB T$ scaling and Planckian behavior ($\nu=1$) and another one with $\nu^*<1$ at higher frequency, most apparent on the optical conductivity itself in contrast to $1/\tau$. A number of theoretical approaches have considered a power-law dependence of the conductivity \cite{Hartnoll-2010, Meyer-2011, Chubukov-2014, Horowitz-2013, Donos-2014, Kiritsis-2015, Rangamani-2015, Langley-2015, Limtragool-2018, LaNave-2019} without resolving this puzzle. A notable exception is the work of Norman and Chubukov \cite{Norman-2006}. The basic assumption of this work is that the electrons are coupled to a Marginal Fermi Liquid susceptibility \cite{Varma-1989, Littlewood-1991, Varma-2002, Varma-2020}. The logarithmic behavior of the susceptibility and corresponding high-energy cut-off observed to be $\sim 0.4$~eV with ARPES \cite{Chang-2008}, is responsible for the apparent sub-linear power law behavior of the optical conductivity. Our work broadens and amplifies this observation. A quantitative description of all aspects at low and high energy in one fell swoop has, to the best of our knowledge, not been presented to this day.

Here we present systematic measurements of the optical spectra, as well as dc resistivity, of a La$_{2-x}$Sr$_x$CuO$_4$ (LSCO) sample with $x=p=0.24$ close to the pseudogap critical point, over a broad range of temperature and frequency. We demonstrate that the data display Planckian quantum critical scaling over an unprecedented range of $\hbar\omega/\kB T$. Furthermore, a direct analysis of the data reveals a logarithmic temperature dependence of the optical effective mass. This establishes a direct connection to another hallmark of Planckian behavior, namely the logarithmic enhancement of the specific heat coefficient $C/T\sim\ln T$ previously observed for LSCO at $p=0.24$ \cite{Girod-2021} as well as for other cuprate superconductors such as Eu-LSCO and Nd-LSCO \cite{Michon-2019}.

We introduce a theoretical framework which relies on a minimal Planckian scaling \textit{Ansatz} for the inelastic scattering rate. We show that this provides an excellent description of the experimental data. Our theoretical analysis offers, notably, a solution to the puzzle mentioned above. Indeed we show that, despite the purely Planckian \textit{Ansatz} which underlies our model, the optical conductivity computed in this framework is well described by an apparent power law with $\nu^*<1$ over an intermediate frequency regime, as also observed in our experimental data. The effective exponent $\nu^*$ is found to be non-universal and to depend on the inelastic coupling constant, which we determine from several independent considerations. The proposed theoretical analysis provides a unifying framework in which the behavior of the $T$-linear resistivity, $\ln T$ behavior of $C/T$, and scaling properties of the optical spectra can all be understood in a consistent manner.

\section{Results}

\subsection{Optical spectra and resistivity}

We measured the optical properties and extracted the complex optical conductivity $\sigma(\omega,T)$ of an LSCO single crystal with a-b orientation (CuO$_2$ planes). The hole doping is $p=x=0.24$, which places our sample above and close to the pseudogap critical point of the LSCO family \cite{Laliberte-2016, Boebinger-1996, Cooper-2009}. The pseudogap state for $T<T^*$, $p<p^*$ is well characterized by transport measurements \cite{Cyr-Choiniere-2018} and ARPES \cite{Matt-2015}. The relatively low $T_c=19$~K of this sample is interesting for extracting the normal-state properties in optics down to low temperatures without using any external magnetic field. In particular, this sample is the same LSCO $p=0.24$ sample as in Ref.~\onlinecite{Michon-2021}, where the evolution of optical spectral weights as a function of doping was reported.

\begin{figure}[tb]
\includegraphics[width=\columnwidth]{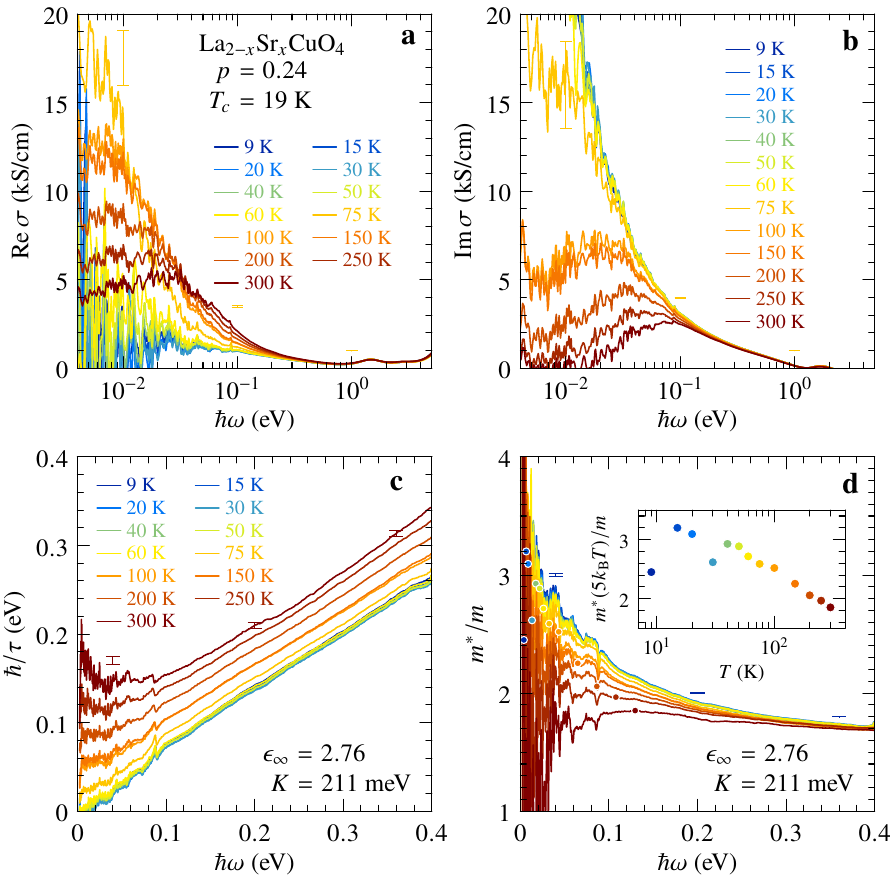}
\caption{\label{fig:fig1}
\textbf{\boldmath Optical data of La$_{2-x}$Sr$_x$CuO$_4$ at $p=0.24$.}
\textbf{a} Real and \textbf{b} imaginary part of the optical conductivity $\sigma$ deduced from the dielectric function $\epsilon$ (\EDF~\ref{fig:fig6}), using Eq.~(\ref{eq:sigma-from-eps}) and the value $\epsilon_{\infty}=2.76$. \textbf{c} Scattering rate and \textbf{d} effective mass deduced from Eqs.~(\ref{eq:tau}) and (\ref{eq:m*}) using $K=211$~meV. The values of $\epsilon_{\infty}$ and $K$ are discussed and justified in the text. Inset: Temperature dependence of $m^*/m$ at $\hbar\omega=5\kB T$ (see dots in \textbf{d}). In each panel errorbars are indicated for three representative frequencies and pertain to the upper curve, \textit{i.e.}, the lowest temperature for $\sigma(\omega)$, $m^*(\omega)/m$ and the highest temperature for $\hbar/\tau(\omega)$. They represent the uncertainty arising from reflectivity calibration using in-situ gold evaporation, and have been estimated by repeating the Kramers--Kronig analysis after multiplying the reflectivity curves by $1 \pm 0.002$.
}
\end{figure}

The quantity probed by the optical experiments of the present study is the planar complex dielectric function $\epsilon(\omega)$. The dielectric function has contributions from the free charge carriers, as well as interband (bound charge) contributions. In the limit $\omega\rightarrow 0$, the latter contribution converges to a constant real value, traditionally indicated with the symbol $\epsilon_{\infty}$:
\begin{align}
    \label{eq:dielectric_function}
    \epsilon(\omega)&=\epsilon_{\infty}+i\frac{\sigma(\omega) }{\epsilon_0 \omega}\\
    \label{eq:free_carrier}
    \sigma(\omega)&=i\frac{e^2K /(\hbar d_c)}{\hbar\omega+M(\omega)}.
\end{align}
Here the free-carrier response $\sigma(\omega)$ is given by the generalized Drude formula, where all dynamical mass renormalization ($m^*/m$) and relaxation ($\hbar/\tau$) processes are represented by a memory-function \cite{Gotze-1972, Basov-2011} 
\begin{equation}\label{eq:memory}
    M(\omega)=\hbar\omega\left[\frac{m^*(\omega)}{m}-1\right]+i\frac{\hbar}{\tau(\omega)}.
\end{equation}
The free-carrier spectral weight per plane is given by the constant $K$ and the interplanar spacing is $d_c$. The scattering rate $\hbar/\tau(\omega)$ deduced using Eqs.~(\ref{eq:dielectric_function}, \ref{eq:free_carrier}, \ref{eq:memory}) and the values of $K$ and $\epsilon_{\infty}$ discussed below are displayed in Fig.~\ref{fig:fig1}\textbf{c}. It depends linearly on frequency for $\kB T\ll\hbar\omega\lesssim0.4$~eV and approaches a constant value for $\hbar\omega<\kB T$. This behavior is similar to that reported for Bi2212 \cite{vanderMarel-2003}. The sign of the curvature above 0.4~eV depends on $\epsilon_{\infty}$ and changes from positive to negative near $\epsilon_{\infty}=4.5$. Our determination $\epsilon_{\infty}=2.76$ presented in Sec.~\ref{sec:scaling} does not take into account data for $\hbar\omega>0.4$~eV and may therefore yield unreliable values of $\hbar/\tau$ in that range (see \SIS~\ref{app:epsinv} and \ref{app:scaling}). 

This linear dependence of the scattering rate calls for a comparison with resistivity. Hence we have also measured the temperature dependence of the resistivity of our sample under two magnetic fields $H=0$~T and $H=16$~T. As displayed in Fig.~\ref{fig:fig2}\textbf{a}, the resistivity has a linear $T$-dependence $\rho=\rho_0+A T$ over an extended range of temperature, with $A\approx 0.63\,\mu\Omega~\mathrm{cm/K}$. This is a hallmark of cuprates in this regime of doping \cite{Daou-2009, Cooper-2009, Legros-2019, Lizaire-2021, Giraldo-Gallo-2018}. It is qualitatively consistent with the observed linear frequency dependence of the scattering rate and, as discussed later in this paper, also in good quantitative agreement with the $\omega\rightarrow 0$ extrapolation of our optical data within experimental uncertainties.

The optical mass enhancement $m^*(\omega)/m$ is displayed in Fig.~\ref{fig:fig1}\textbf{d}. With the chosen normalization, $m^*/m$ does not reach the asymptotic value of one in the range $\hbar\omega<0.4$~eV, which means that intra- and interband and/or mid-infrared transitions overlap above 0.4~eV. The inset of Fig.~\ref{fig:fig1}\textbf{d} shows a semi-log plot of the mass enhancement evaluated at $\hbar\omega=5\kB T$, where the noise level is low for $T\geqslant 40$~K. Despite the larger uncertainties at low $T$, this plot clearly reveals a logarithmic temperature dependence of $m^*/m$. This is a robust feature of the data, independent of the choice of $\epsilon_{\infty}$ and $K$. We note that the specific heat coefficient $C/T$ of LSCO at the same doping level was previously reported to display a logarithmic dependence on temperature, see Fig.~\ref{fig:fig2}\textbf{c} \cite{Michon-2019, Girod-2021}. We will further elaborate on this important finding of a logarithmic dependence of the optical mass and discuss its relation to specific heat in the next section.

\subsection{Scaling analysis}
\label{sec:scaling}

In this section, we consider simultaneously the frequency and temperature dependence of the optical properties and investigate whether $\hbar\omega/\kB T$ scaling holds for this sample close to the pseudogap critical point. We propose a procedure to determine the three parameters $\epsilon_\infty$, $K$, and $m$ introduced above.

\subsubsection{Putting $\omega/T$ scaling to the test}

Quantum systems close to a quantum critical point display scale invariance. Temperature being the only relevant energy scale in the quantum critical regime, this leads in many cases to $\omega/T$ scaling \cite{Sachdev-2011} (in most of the discussion below, we set $\hbar=\kB=1$ except when mentioned explicitly). In such a system we expect the complex optical conductivity to obey a scaling behavior $1/\sigma(\omega,T)\propto T^{\nu}F(\omega/T)$, with $\nu\leqslant 1$ a critical exponent. More precisely, the scaling properties of the optical scattering rate and effective mass read: 
	\begin{align}
		1/\tau(\omega,T)&= T^{\nu}f_{\tau}(\omega/T)\\
		\label{eq:scaling_mstar}
		m^*(\omega,T)-m^*(0,T)&= T^{\nu-1}f_m(\omega/T)
	\end{align}
with $f_\tau$ and $f_m$ two scaling functions. This behavior requires that both $\hbar\omega$ and $\kB T$ are smaller than a high-energy electronic cutoff, but their ratio can be arbitrary. Furthermore, we note that when $\nu=1$ (Planckian case) the scaling is violated by logarithmic terms, which control in particular the zero-frequency value of the optical mass $m^*(0,T)$. As shown in Sec.~\ref{sec:theory} within a simple theoretical model, $\omega/T$ scaling nonetheless holds in this case to an excellent approximation provided that $m^*(0,T)$ is subtracted, as in Eq.~(\ref{eq:scaling_mstar}). We also note that in a Fermi liquid, the single-particle scattering rate $\propto \omega^2 + (\pi T)^2$ does obey $\omega/T$ scaling (with formally $\nu=2$), but the optical conductivity does not. Indeed, it involves $\omega/T^2$ terms violating scaling, and hence depends on two scaling variables $\omega/T^2$ and $\omega/T$, as is already clear from an (approximate) generalized Drude expression $1/\sigma \approx -i\omega + \tau_0[\omega^2+(2\pi T)^2]$. For a detailed discussion of this point, see Ref.~\onlinecite{Berthod-2013}. Such violations of scaling by $\omega/T^\nu$ terms apply more generally to the case where the scattering rate varies as $T^\nu$ with $\nu>1$. Hence, $\omega/T$ scaling for both the optical scattering rate and optical effective mass are a hallmark of non-Fermi liquid behavior with $\nu\leqslant 1$. Previous work has indeed provided evidence for $\omega/T$ scaling in the optical properties of cuprates \cite{vanderMarel-2003, vanderMarel-2006}.

\begin{figure}[tb]
\includegraphics[width=\columnwidth]{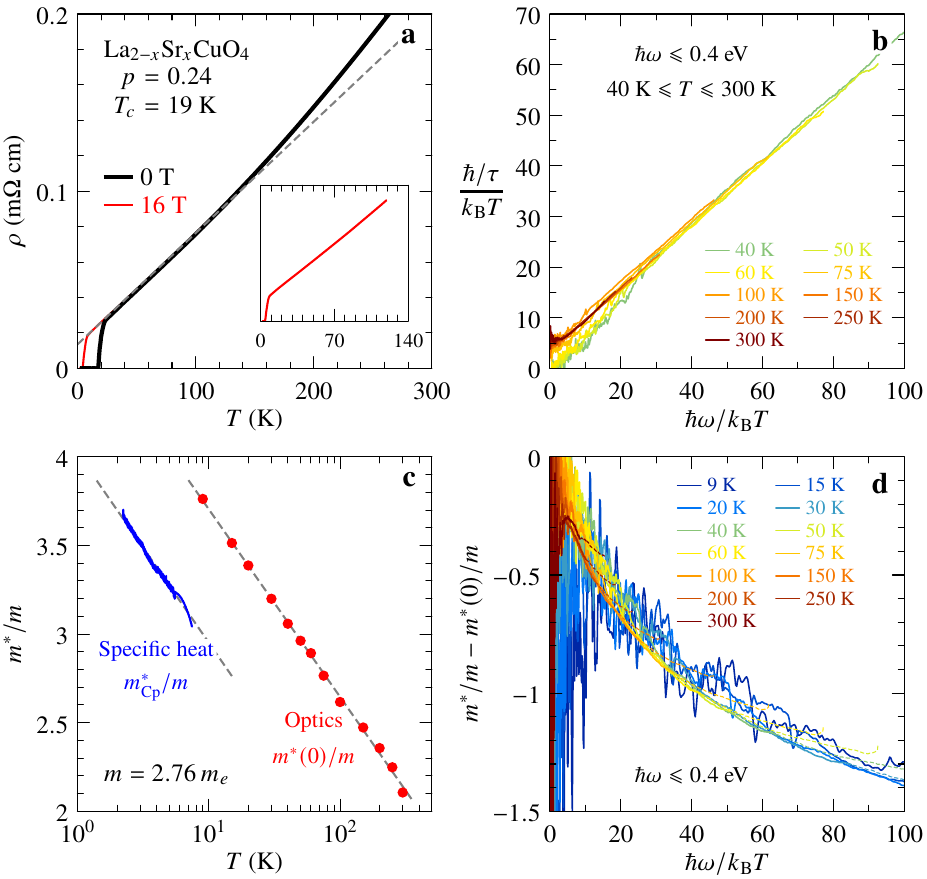}
\caption{\label{fig:fig2}
\textbf{Scaling of scattering rate and mass enhancement.}
\textbf{a} Temperature-dependent resistivity measured in zero field (black) and at 16~teslas (red). The inset emphasizes the linearity of the 16~T data at low temperature. The dashed line shows $\rho_0+AT$ with $\rho_0=12.2~\mu\Omega~\mathrm{cm}$ and $A=0.63~\mu\Omega~\mathrm{cm/K}$. \textbf{b} Scattering rate divided by temperature plotted versus $\omega/T$; the collapse of the curves indicates a behavior $1/\tau\sim Tf_{\tau}(\omega/T)$. \textbf{c} Effective quasiparticle mass (in units of the indicated band mass $m$) deduced from the low-temperature electronic specific heat \cite{Girod-2021} [$m^*_{\mathrm{Cp}}=(3/\pi)(\hbar^2d_c/\kB^2)(C/T)$] and zero-frequency optical mass enhancement; the dashed lines indicate $\ln T$ behavior. \textbf{d} Optical mass minus the zero-frequency mass shown in \textbf{c} plotted versus $\omega/T$; the collapse of the curves indicates a behavior $m^*(\omega)-m^*(0)\sim f_m(\omega/T)$. The data between 0.22 and 0.4~eV are shown as dotted lines. $\epsilon_{\infty}=2.76$ was used here as in Fig.~\ref{fig:fig1}.
}
\end{figure}

Here, we investigate whether our optical data obey $\omega/T$ scaling. We find that the quality of the scaling depends sensitively on the chosen value of $\epsilon_\infty$. Different prescriptions in the literature to fix $\epsilon_{\infty}$ yield~---~independently of the method used~---~values ranging from $\epsilon_\infty \approx 4.3$ for strongly underdoped Bi2212 to $\epsilon_\infty \approx 5.6$ for strongly overdoped Bi2212 \cite{vanHeumen-2007, Hwang-2007}. The parameter $\epsilon_{\infty}$ is commonly understood to represent the dielectric constant of the material in the absence of the charge carriers, and is caused by the bound charge responsible for interband transitions at energies typically above 1~eV. While this definition is unambiguous for the insulating parent compound, for the doped material one is confronted with the difficulty that the optical conductivity at these higher energies also contains contributions described by the self-energy of the conduction electrons, caused for example by their coupling to dd-excitations \cite{Barantani-2022}. Consequently, not all of the oscillator strength in the interband region represents bound charge. Our model overcomes this hurdle by determining the low-energy spectrum below 0.4~eV, and subsuming all bound charge contributions in a single constant $\epsilon_{\infty}$. Its value is expected to be bound from above by the value of the insulating phase, in other words we expect to find $\epsilon_{\infty}<4.5$ (see \SIS~\ref{app:epsinv}). Rather than setting an \textit{a priori} value for $\epsilon_{\infty}$, we follow here a different route and we choose the value that yields the best scaling collapse for a given value of the exponent $\nu$. This program is straightforwardly implemented for $1/\tau$ and indicates that the best scaling collapse is achieved with $\nu\approx1$ and $\epsilon_{\infty}\approx 3$, see Fig.~\ref{fig:fig2}\textbf{b} as well as \SIS~\ref{app:scaling} and \EDF~\ref{fig:fig7}. Turning to $m^*$, we found that subtracting the dc value $m^*(\omega=0,T)$ is crucial when attempting to collapse the data. Extrapolating optical data to zero frequency is hampered by noise. Hence, instead of attempting an extrapolation, we consider $m^*(0,T)$ as adjustable values that we again tune such as to optimize the collapse of the optical data. This analysis of $m^*/m$ confirms that the best scaling collapse occurs for $\nu\approx 1$ but indicates a larger $\epsilon_{\infty}\approx 7$ (\SIS~\ref{app:scaling} and \EDF~\ref{fig:fig8}). The determination of $\epsilon_{\infty}$ from the mass data depends sensitively on the frequency range tested for scaling and drops to value below $\epsilon_{\infty}=3$ when focusing on lower frequencies. As a third step, we perform a simultaneous optimization of the data collapse for $1/\tau$ and $m^*/m$, which yields the values $\nu=1$, $\epsilon_{\infty}=2.76$ which we will adopt throughout the following. Note that a determination of $\epsilon_{\infty}$ by separation of the high-frequency modes in a Drude--Lorentz representation of $\epsilon(\omega)$ yields a larger value $\epsilon_{\infty}=4.5\pm0.5$, as typically found in the cuprates \cite{vanderMarel-2003, Carbone-2006, Hwang-2007}. Importantly, all our conclusions hold if we use this latter value in the analysis, however the quality of the scaling displayed in Figs.~\ref{fig:fig2} and \ref{fig:fig5} is slightly degraded.

\subsubsection{Scaling of the optical scattering rate and connection to resistivity}

The scaling properties of the scattering rate obtained from our optical data according to the procedure described above is illustrated in Fig.~\ref{fig:fig2}\textbf{b}, which displays $\hbar/\tau$ divided by $\kB T$ and plotted versus $\hbar\omega/\kB T$ for temperatures above the superconducting transition. The collapse of the curves at different temperatures reveals the behavior $\hbar/\tau\propto Tf_{\tau}(\omega/T)$. The function $f_{\tau}(x)$ reaches a constant $f_{\tau}(0)>0$ at small values of the argument, and behaves for large arguments as $f_{\tau}(x\gg1)\propto x$. This is consistent with the typical quantum critical behavior $\hbar/\tau\sim\max(T,\omega)$. When inserted in the $\omega=0$ limit of Eq.~(\ref{eq:tau-and-m*}), the value $f_{\tau}(0)\approx5$ indicated by Fig.~\ref{fig:fig2}\textbf{b} yields $1/\sigma(0)=AT$ with $A=0.55~\mu\Omega~\mathrm{cm/K}$, in fairly good agreement with the measured resistivity (Fig.~\ref{fig:fig2}\textbf{a}). Hence the resistivity and optical-spectroscopy data are fully consistent, both of them supporting a Planckian dissipation scenario with $\nu=1$ for LSCO at $p=0.24$.

\subsubsection{Spectral weight, effective mass and connection to specific heat}

The dc mass enhancement values $m^*(0,T)/m$ resulting from the procedure described above are displayed in Fig.~\ref{fig:fig2}\textbf{c}. Remarkably, as seen on this figure, the scaling analysis delivers an almost perfectly logarithmic temperature dependence of $m^*(0,T)$, consistent with a Planckian behavior $\nu=1$. As mentioned above, this logarithmic behavior can actually be identified in the unprocessed optical data, (see inset of Fig.~\ref{fig:fig1}). In order to compare this behavior to the corresponding logarithmic behavior reported for the specific heat, we note that the scaling analysis provides $m^*(0,T)$ up to a multiplicative constant $Km$, where $m$ is the band mass. In contrast, the electronic specific heat yields the quasiparticle mass in units of the bare electron mass $m_e$. We expect that the logarithmic $T$-variation of $m^*(0,T)$ and $m^*_{\mathrm{qp}}\propto C/T$ are both due to the critical inelastic scattering and that the $\ln T$ term in each quantity should therefore have identical prefactors. Imposing this identity provides a relationship between $Km$ and $m_e$, namely $(m/m_e)K=583$~meV. 

Remarkably, we have found that this condition is obeyed within less than a percent by a square-lattice tight-binding model with parameters appropriate for LSCO at $p=0.24$ (\SIS~\ref{app:tight-binding}). This model has nearest and next-nearest neighbor hopping amplitudes $t=0.3$~eV and $t'/t=-0.17$ \cite{Pavarini-2001}, respectively, and an electronic density $n=0.76/a^2$. The Fermi-level density of states is $1.646/(\mathrm{eV}a^2)$, which corresponds to a band mass $m/m_e=2.76$ using the LSCO lattice parameter $a=3.78$~\AA{}. The spectral weight is $K=211$~meV, such that the prediction of this tight-binding model is $(m/m_e)K=582$~meV, in perfect agreement with the previously determined value. In view of this agreement, we use the tight-binding model in order to fix the remaining two system parameters: $m=2.76\,m_e$ and $K=211$~meV.

Figure~\ref{fig:fig2}\textbf{c} compares the mass enhancement inferred from the low-temperature specific heat and from the scaling analysis of the optical data. The tight-binding value of the product $Km$ ensures that both data sets have the same slope on a semi-log plot. However, the resulting optical mass enhancement is larger than the quasiparticle mass enhancement by $\approx 0.75$, which is also the amount by which the infrared mass enhancement exceeds unity in Fig.~\ref{fig:fig1}\textbf{d}. A mass enhancement larger than unity at 0.4~eV implies that part of the intraband spectral weight lies above 0.4~eV, overlapping with the interband transitions. Conversely, interband spectral weight is likely leaking below 0.4~eV, which prevents us from accessing the absolute value of the genuine intraband mass by optical means. Figure~\ref{fig:fig2}\textbf{d} shows the collapse of the frequency-dependent change of the mass enhancement, confirming the behavior $m^*(\omega)-m^*(0)\approx T^{\nu-1}f_m(\omega/T)$ with $\nu=1$. The shape of the scaling function $f_m(x)$ agrees remarkably well with the theoretical prediction derived in Sec.~\ref{sec:theory} below.

\subsubsection{Apparent power-law behavior: a puzzle}

The above scaling analysis has led us to the following conclusions. (i) The optical scattering rate and optical mass enhancement of LSCO at $p=0.24$ exhibit $\omega/T$ scaling over two decades for the chosen value $\epsilon_{\infty}=2.76$. (ii) The best collapse of the data is achieved for an exponent $\nu=1$ corresponding to Planckian dissipation. This behavior is consistent with the measured $T$-linear resistivity. (iii) The temperature dependence of $m^*(0,T)$ that produces the best data collapse is logarithmic, consistently with the temperature dependence of the electronic specific heat.

\begin{figure}[tb]
\includegraphics[width=\columnwidth]{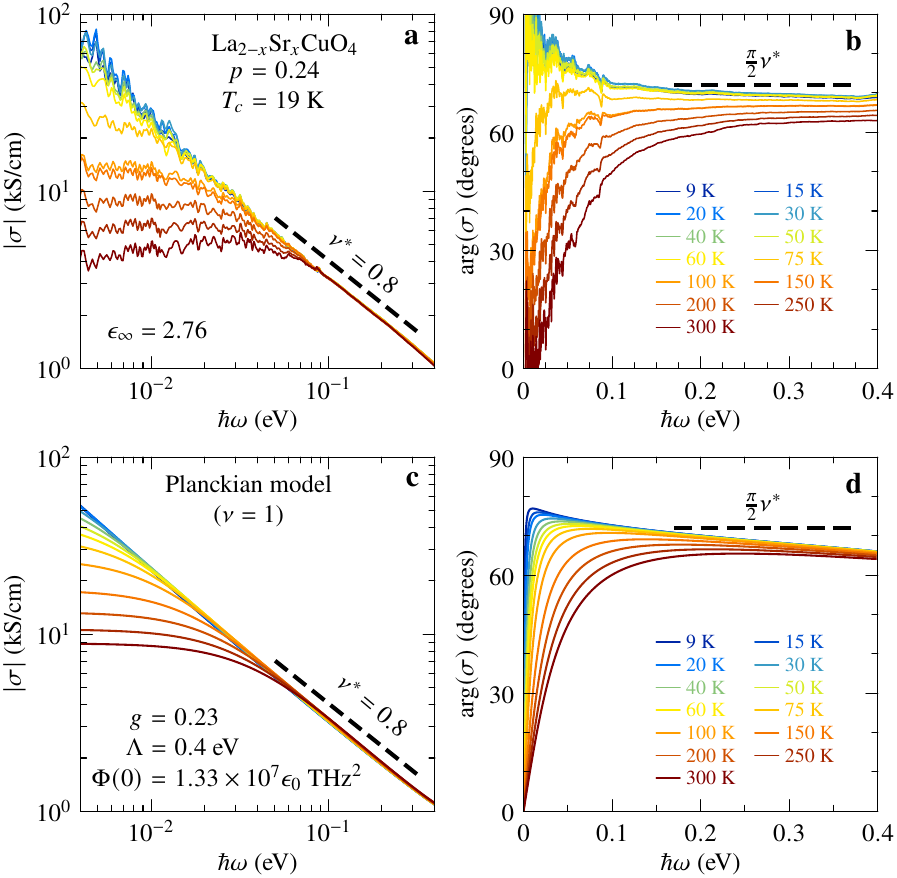}
\caption{\label{fig:fig3}
\textbf{Sub-linear power law at intermediate frequencies.}
\textbf{a} Modulus and \textbf{b} phase of the complex conductivity shown in Figs.~\ref{fig:fig1}\textbf{a} and \ref{fig:fig1}\textbf{b}; the modulus decays with an exponent $\nu^*\approx0.8$ and the phase approaches a value slightly lower than $(\pi/2)\nu^*$. \textbf{c} and \textbf{d}: same quantities calculated using a Planckian model with linear-in-energy scattering rate, Eqs.~(\ref{eq:Sigma}) and (\ref{eq:sigma}). The model and parameters are discussed in the text.
}
\end{figure}

Hence, the data presented in Fig.~\ref{fig:fig2} provide compelling evidence that the low-energy carriers in LSCO at the doping $p=0.24$ experience linear-in-energy and linear-in-temperature inelastic scattering processes, as expected in a scale-invariant quantum critical system characterized by Planckian dissipation. It is therefore at first sight surprising that the infrared conductivity exhibits as a function of frequency a power law with an exponent that is clearly smaller than unity, as highlighted in Figs.~\ref{fig:fig3}\textbf{a} and \ref{fig:fig3}\textbf{b}. These figures show that the modulus and phase of $\sigma$ are both to a good accuracy consistent with the behavior $\sigma\propto(-i\omega)^{-\nu^*}=\omega^{-\nu^*}e^{i\frac{\pi}{2}\nu^*}$ with an exponent $\nu^*=0.8$. A similar behavior with exponent $\nu^*\approx0.6$ was reported for optimally- and overdoped Bi2212 \cite{vanderMarel-2003}, while earlier optical investigations of YBCO and Bi2212 have also reported power law behavior of $\mathrm{Re}\,\sigma(\omega)$ \cite{Schlesinger-1990b, ElAzrak-1994, Baraduc-1996}. We now address this question by considering a theoretical model presented in the following section. As derived there, and illustrated in Figs.~\ref{fig:fig3}\textbf{c} and \ref{fig:fig3}\textbf{d}, we show that an apparent exponent $\nu^*<1$ is actually \emph{predicted by theory} for Planckian systems with single-particle self-energy exponent $\nu=1$, over an intermediate range of values of $\omega/T$. This is one of the central claims of our work.

\subsection{Theory}
\label{sec:theory}

In this section, we consider a simple theoretical model and explore its implications for the optical conductivity. Our central assumption is that the inelastic scattering rate (imaginary part of the self-energy) obeys the following scaling property:
	\begin{equation}\label{eq:Sigma_scaling}
		-\mathrm{Im}\,\Sigma(\varepsilon)=g\pi\kB TS\left(\frac{\varepsilon}{\kB T}\right).
	\end{equation}
In this expression $g$ is a dimensionless inelastic coupling constant and $\varepsilon=\hbar\omega$. This $\hbar\omega/\kB T$ scaling form is assumed to apply when both $\hbar\omega$ and $\kB T$ are smaller than a high-energy cutoff $\Lambda$ but their ratio can be arbitrary. The detailed form of the scaling function $S$ is not essential, except for the requirements that $S(0)$ is finite and $S(x\gg 1)\propto |x|$. These properties ensure that the low-frequency inelastic scattering rate depends linearly on $T$ for $\hbar\omega\ll\kB T$ and that dissipation is linear in energy for $\hbar\omega\gg\kB T$, which are hallmarks of Planckian behavior. We note that such a scaling form appears in the context of microscopic models such as overscreened non-Fermi liquid Kondo models \cite{Parcollet-1998} and the doped SYK model close to a quantum critical point \cite{Sachdev-1993, Kitaev-2015, Parcollet-1999, Dumitrescu-2022, Patel-2022}. In such models, conformal invariance applies and dictates the form of the scaling function to be $S(x)=x\coth(x/2)$ (with possible modifications accounting for a particle-hole spectral asymmetry parameter, see Refs.~\onlinecite{Parcollet-1998, Georges-2021} and \SIS~\ref{app:ph-asymmetry}. We have assumed that the inelastic scattering rate is momentum independent (spatially local) i.e.\ uniform along the Fermi surface. This assumption is supported by recent angular-dependent magnetoresistance experiments on Nd-LSCO at a doping close to the pseudogap quantum critical point \cite{Grissonnanche-2021}~---~see also Ref.~\onlinecite{Millis-2003}. In contrast, the elastic part of the scattering rate (not included in our theoretical model) was found to be strongly anisotropic (angular dependent).

The real part of the self-energy is obtained from the Kramers--Kronig relation which reads, substituting the scaling form above: 
	\begin{equation}\label{eq:Sigma}
		\Sigma(z)=g \kB T\int_{\Lambda}dx\,\frac{S(x)}{z/\kB T-x}.
	\end{equation}
We note that this expression is only defined provided the integral is bounded at high-frequency by the cutoff $\Lambda$, as detailed in \SIS~\ref{app:nu=1}. This reflects into a logarithmic temperature dependence at low energy:
	\begin{equation}
			\mathrm{Re}\,[\Sigma(\varepsilon)-\Sigma(0)]=-2g\varepsilon\ln(a\Lambda/\kB T)
	\end{equation}
with $a=0.770542$ a numerical constant (\SIS~\ref{app:nu=1}). Correspondingly, the effective mass of quasiparticles, as well as the specific heat, is logarithmically divergent at low temperature: 
	\begin{equation}\label{eq:mass}
		\frac{m^*_{\mathrm{qp}}}{m}=\frac{1}{Z}=
		1+2g\ln\left(a\frac{\Lambda}{\kB T}\right)
	\end{equation}
with $1/Z=1-d\mathrm{Re}\,\Sigma(\varepsilon)/d\varepsilon|_{\varepsilon=0}$. Importantly, the coefficient of the dominant $\ln T$ term depends only on the value of the inelastic coupling $g$.

In a local (momentum-independent) theory, vertex corrections are absent \cite{Khurana-1990, Vucicevic-2021} and the optical conductivity can thus be directly computed from the knowledge of the self-energy as \cite{Allen-2015}: 
	\begin{equation}\label{eq:sigma}
		\sigma(\omega)=\frac{i\Phi(0)}{\omega}\int_{-\infty}^{\infty}d\varepsilon\,
		\frac{f(\varepsilon)-f(\varepsilon+\hbar\omega)}
		{\hbar\omega+\Sigma^*(\varepsilon)-\Sigma(\varepsilon+\hbar\omega)}
	\end{equation}
where $f(\varepsilon)=(e^{\varepsilon/\kB T}+1)^{-1}$ is the Fermi function and $\Sigma^*$ denotes complex conjugation. In this expression
	$\Phi(\varepsilon)=2(e/\hbar)^2\int_{\mathrm{BZ}}\frac{d^2k}{(2\pi)^2}
	\left(\partial\varepsilon_{\vec{k}}/\partial k_x\right)^2
	\delta(\varepsilon+\mu_0-\varepsilon_{\vec{k}})$
is the transport function associated with the bare bandstructure. We have assumed that its energy dependence can be neglected so that only the value $\Phi(0)$ at the Fermi level matters (we set $\mu_0=0$ by convention). Using a tight-binding model for the band dispersion, $\Phi(0)$ can be related to the spectral weight $K$ discussed in the previous section as: $(\hbar/e)^2\Phi_{\mathrm{2D}}(0)=K=211$~meV, i.e.\ $\Phi(0)=\Phi_{\mathrm{2D}}(0)/d_c=1.33\times10^7\epsilon_0~\mathrm{THz}^2$ (see \SIS~\ref{app:tight-binding}). 

\begin{figure}[tb]
\includegraphics[width=\columnwidth]{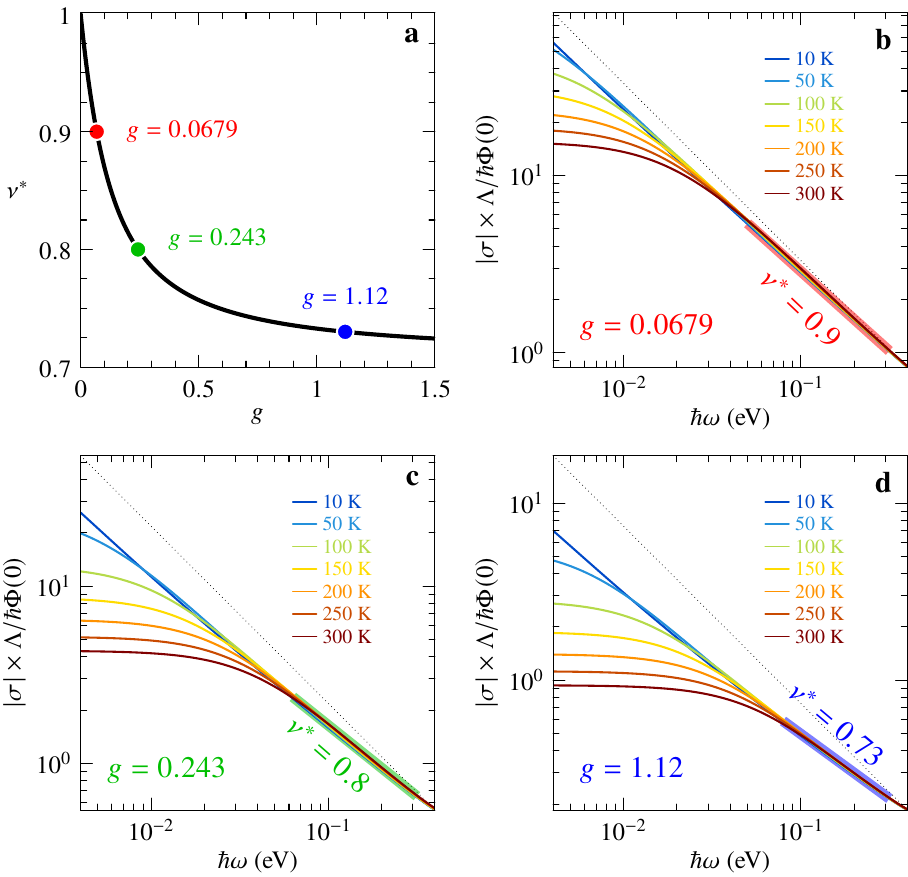}
\caption{\label{fig:fig4}
\textbf{Effective exponent.}
Emergence of an apparent sub-linear power-law in a pure Planckian model. \textbf{a} Apparent exponent given by Eq.~(\ref{eq:nustar}) versus interaction strength $g$. (b--d) Modulus of the optical conductivity on a log-log scale showing the apparent power law at energies between $\kB T$ and the cutoff $\Lambda=0.4$~eV. Data are shown for three values of $g$ (dots in \textbf{a}) and a range of temperatures. Both horizontal and vertical axes cover exactly two decades, such that a $1/\omega$ behavior would correspond to a slope of $-1$ (dotted line).
}
\end{figure}

Within our model, the behavior of the optical conductivity relies on three parameters: the cutoff $\Lambda$, the Drude weight related to $\Phi(0)$ and, importantly, the dimensionless inelastic coupling $g$. An analysis of Eq.~(\ref{eq:sigma}), detailed in \SIS~\ref{app:nu=1}, yields the following behavior in the different frequency regimes:
\begin{itemize}

\item $\hbar\omega\lesssim \kB T$. The optical conductivity in this regime takes a Drude-like form Eq.~(\ref{eq:tau-and-m*}) with $\hbar/\tau=4\pi g\kB T$. The numerically computed zero-frequency optical mass enhancement $m^*(0)/m$ agrees very well with $m^*_{\mathrm{qp}}/m=1/Z$ as given by Eq.~(\ref{eq:mass}), see \EDF~\ref{fig:fig11}. Fitting Eq.~(\ref{eq:mass}) to the $m^*(0)/m$ data in Fig.~\ref{fig:fig2}\textbf{c} provides the values $g=0.23$ and $\Lambda=0.4$~eV. 

\item $\hbar\omega\gtrsim\Lambda$. In this high-frequency regime, the asymptotic behavior is fixed by causality and reads $|\sigma|\sim1/\omega$, $\mathrm{arg}(\sigma)\to \pi/2$ (see \EDF~\ref{fig:fig9} and \EDF~\ref{fig:fig10}).

\item $\kB T\lesssim\hbar\omega\lesssim\Lambda$. In this regime, which is the most important in practice when considering our experimental data, one can derive the following expression: 
	\begin{equation}\label{eq:sigmaII}
		\sigma(\omega)\approx\frac{\Phi(0)}{-i\omega}\frac{1}{1
		+2g\left[1-\ln\left(\frac{\hbar\omega}{2\Lambda}\right)\right]
		+i\pi g}.
	\end{equation}
Remarkably, as shown in Fig.~\ref{fig:fig4}, the theoretical optical conductivity is very well approximated in this regime by an apparent power-law dependence $|\sigma | \sim |\omega |^{-\nu^*}$, over at least a decade in frequency. The effective exponent $\nu^*<1$ depends continuously on the inelastic coupling constant $g$ and can be estimated as:
	\begin{align}
		\nonumber
		\nu^* &\equiv -\left.\frac{d\ln|\sigma|}{d\ln\omega}\right|_{\hbar\omega=\Lambda/2}\\
		\label{eq:nustar}
		&=1-\frac{2g[1+2g(1+\ln4)]}{\pi^2g^2+[1+2g(1+\ln4)]^2}.
	\end{align}
Correspondingly, $\mathrm{arg}(\sigma)$ has a plateau at $\mathrm{arg}(\sigma)\approx\pi\nu^*/2$ before reaching its eventual asymptotic value $\pi/2$ (\EDF~\ref{fig:fig10}). Using the value $g=0.23$ deduced above from $m^*(0)/m$ yields $\nu^*=0.8$, in very good agreement with experiment, as shown in Fig.~\ref{fig:fig3}.

\end{itemize}

In the dc limit $\omega\rightarrow 0$, Eq.~(\ref{eq:sigma}) together with our \textit{Ansatz} for the scattering rate, yields a $T$-linear resistivity: 
	\begin{equation}\label{eq:A}
	\rho=AT,\quad
		A=\frac{4\pi^3 \kB }{7\zeta(3)\hbar}\frac{g}{\Phi(0)}
		=\frac{4\pi^3\hbar \kB d_c}{7\zeta(3)e^2}\frac{g}{K}.
	\end{equation}
Using the values of $g$ and $\Phi(0)$ determined above, we obtain: $A=0.38~\mu\Omega~\mathrm{cm/K}$, to be compared to the experimental value $A = 0.63~\mu\Omega~\mathrm{cm/K}$. It is reassuring that a reasonable order of magnitude is obtained (at the 60\% level) for the $A$-coefficient, while obviously a precise quantitative agreement cannot be expected from such a simple model. 

\begin{figure}[tb]
\includegraphics[width=\columnwidth]{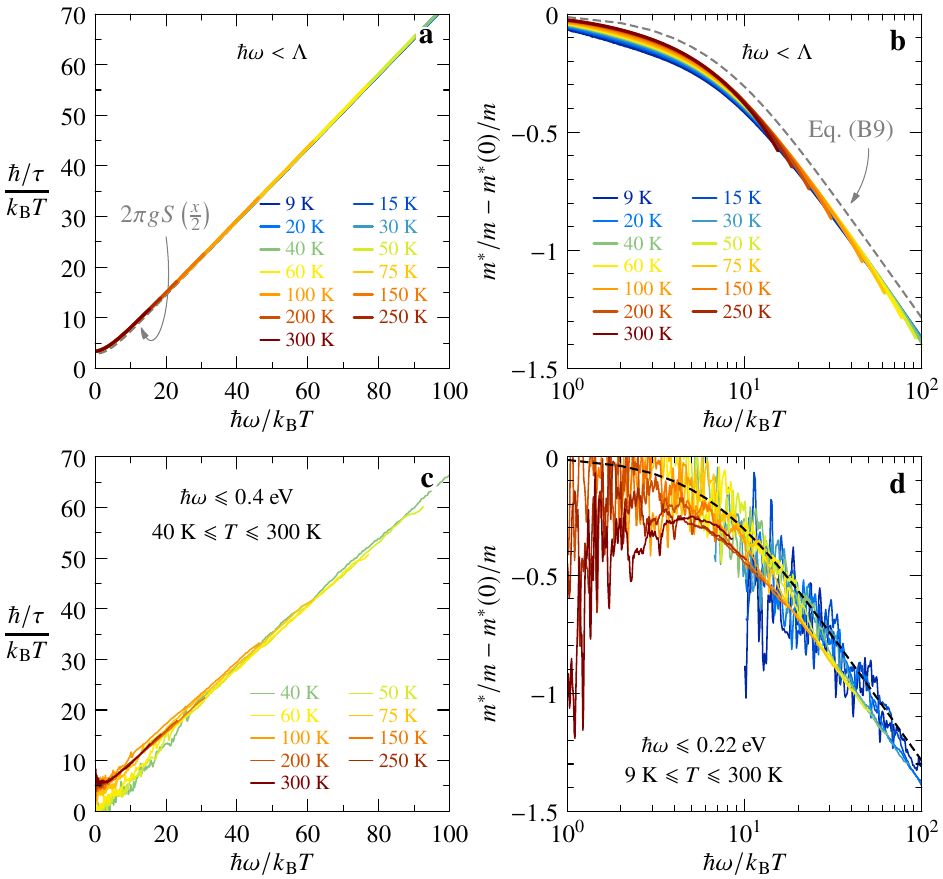}
\caption{\label{fig:fig5}
\textbf{Frequency-temperature scaling.}
\textbf{a} Approximate collapse of the theoretical scattering rate and \textbf{b} mass enhancement; the dashed lines show $2\pi gS(x/2)$ in \textbf{a} and Eq.~(\ref{eq:fm}) in \textbf{b}. \textbf{c} Same data as in Fig.~\ref{fig:fig2}\textbf{b}. \textbf{d} Same data as in Fig.~\ref{fig:fig2}\textbf{d} on a logarithmic scale (not displayed here because of excessive noise: $\hbar\omega/\kB T<10$ for $T<T_c$); the dashed line is Eq.~(\ref{eq:fm}).
}
\end{figure}

Finally, we present in Fig.~\ref{fig:fig5} an $\omega/T$ scaling plot of $1/\tau$ and $m^*/m-m^*(0)/m$ for our model, as well as a direct comparison to experimental data. We emphasize that $\omega/T$ scaling does not hold exactly for either of these quantities within our Planckian model. This is due to the fact that the real part of the self-energy behaves logarithmically at low $T$ and thus leads to violations of scaling, as also clear from the need to retain a finite cutoff $\Lambda$. However, approximate $\omega/T$ scaling is obeyed to a rather high accuracy, as shown in panels \textbf{a} and \textbf{b} of Fig.~\ref{fig:fig5} and discussed in more details analytically in \SIS~\ref{app:nu=1}. Panels \textbf{c} and \textbf{d} allow for a direct comparison between the scaling properties of the theoretical model and the experimental data, including analytical expressions of the approximate scaling functions derived in \SIS~\ref{app:nu=1}. These functions stem from an approximate expression for the conductivity, Eq.~(\ref{eq:sigma23}), that displays exact $\omega/T$ scaling. The approximation made in deriving them explains why the scaling functions differ slightly from the numerical data in Figs.~\ref{fig:fig5}\textbf{a} and \ref{fig:fig5}\textbf{b}. Note the similar difference with the experimental data in Fig.~\ref{fig:fig5}\textbf{d}.

\section{Discussion}

In this article, we have shown that our experimental optical data for LSCO at $p=0.24$ display scaling properties as a function of $\hbar\omega/\kB T$ which are consistent with Planckian behavior corresponding to a scaling exponent $\nu=1$. We found that the accuracy of the data scaling depends on the choice of the parameter $\epsilon_\infty$ relating the optical conductivity to the measured dielectric permittivity, and that optimal scaling is achieved for a specific range of values of this parameter. 

From both a direct analysis of the optical data and by requiring optimal scaling, we demonstrated that the low-frequency optical effective mass $m^*(\omega\approx 0,T)/m$ displays a logarithmic dependence on temperature. This dependence, also a hallmark of Planckian behavior, is qualitatively consistent with that reported for the specific heat (quasiparticle effective mass) \cite{Michon-2019, Girod-2021}. We showed that the coefficient of the logarithmic term can be made quantitatively consistent between these two measurements if a specific relation exists between the spectral weight $K$ and the ratio $m/m_e$ of the band mass to the bare electron mass. Interestingly, we found that a realistic tight-binding model satisfies this relation. The low-frequency optical scattering rate $1/\tau$ extracted from our scaling analysis displays a linear dependence on temperature, consistent with the $T$-linear dependence of the resistivity that we measured on the same sample, with a quite good quantitative agreement found between the $T$-linear slopes of these two measurements. 

We have introduced a simple theoretical model which relies on the assumption that the single-particle inelastic scattering rate (imaginary part of the self-energy) displays $\hbar\omega/\kB T$ scaling properties with $\nu=1$ and that its angular dependence along the Fermi surface can be neglected. These assumptions are consistent with angular dependent magnetoresistance measurements \cite{Grissonnanche-2021}. The model involves a dimensionless inelastic coupling constant $g$ as a key parameter. We calculated the optical conductivity based on this model and showed that it accounts very well for the frequency dependence (Fig.~\ref{fig:fig3}) and $\omega/T$ scaling properties (Fig.~\ref{fig:fig5}) of our experimental data. 

A key finding of our analysis is that the calculated optical conductivity displays an \emph{apparent} power-law behavior with an effective exponent $\nu^*<1$ over an extended frequency range relevant to experiments (Figs.~\ref{fig:fig3} and \ref{fig:fig4}). We were able to establish that $\nu^*$ depends continuously on the inelastic coupling constant $g$ [Eq.~(\ref{eq:nustar}) and Fig.~\ref{fig:fig4}\textbf{a}]. This apparent power law is also clear in the experimental data, especially when displaying the data for $|\sigma|$ and $\mathrm{arg}(\sigma)$ as a function of frequency. Hence, our analysis solves a long-standing puzzle in the field, namely the seemingly contradictory observations of Planckian behavior with $\nu=1$ for the resistivity and optical scattering rate versus a power law $\nu^*<1$ observed for $|\sigma|$ and $\mathrm{arg}(\sigma)$. We note that the apparent exponent $\nu^*$ reported in previous optical spectroscopy literature varies from one compound to another, which is consistent with our finding that $\nu^*$ depends on $g$ and is hence not universal. For our LSCO sample, the measured value of $\nu^*$ leads to the value $g\approx 0.23$. 

The logarithmic temperature dependence of both the optical effective mass and the quasiparticle effective mass is directly proportional to the inelastic coupling constant $g$. We emphasize that this is profoundly different from what happens in a Fermi liquid. There, using the Kramers--Kronig relation, one sees that the effective mass enhancement (related to the low-frequency behavior of the real part of the self-energy) depends on the whole high-frequency behavior of the imaginary part of the self-energy. In contrast, in a Planckian metal obeying $\omega/T$ scaling, the dominant $\ln T$ dependence of the mass is entirely determined by the low-energy behavior of the imaginary part of the self-energy, see Eq.~(\ref{eq:mass}). Based on this observation, we found that the slope of the $\ln T$ term in the effective mass and specific heat is consistent with the value $g\approx 0.23$ independently determined from the effective exponent $\nu^*$. Using that same value of $g$ within our simple theory leads to a value of the prefactor $A$ of the $T$-linear term in the resistivity which accounts for 60\% of the experimentally measured value. Quantitative agreement would require $g\approx 0.38$, corresponding to a value of $\nu^*\approx 0.77$ also quite close to the experimentally observed value $\nu^*\approx 0.8$. It is also conceivable that electron-phonon coupling contributes to the experimental value of $A$. In view of the extreme simplification of the theoretical model for transport used in the present work, it is satisfying that overall consistency between optics, specific heat and resistivity can be achieved with comparable values of the coupling $g$.

In recent works \cite{Georges-2021, Gourgout-2022}, Planckian behavior has also been put forward as an explanation for the observed unconventional temperature dependence of the in-plane and $c$-axis Seebeck coefficient of Nd-LSCO. In these works, the same scaling form of the inelastic scattering rate than the one used here was used, modified by a particle-hole asymmetry parameter. For simplicity, this asymmetry parameter was set to zero in the present article. We have checked, as detailed in \SIS~\ref{app:ph-asymmetry}, that our results and analysis are unchanged if this asymmetry parameter is included, as is expected from the fact that optical spectroscopy measures particle-hole excitations and is thus rather insensitive to the value of the particle-hole asymmetry parameter. 

Finally, we note for completeness that a power-law behavior of the optical conductivity has also been observed in other materials, including quasi one-dimensional conductors \cite{Schwartz-1998, Pashkin-2006, Lavagnini-2009, Lee-2005} with $\nu^*\sim 1.5$, and three-dimensional conductors \cite{Cao-1997, Kostic-1998, Dodge-2000, Mena-2003} with $\nu^*\sim 0.5$. In the former case, Luttinger-liquid behavior provides an explanation for the observed power law at intermediate frequencies \cite{Schwartz-1998}, while the interpretation of the power-law behavior for materials such as Sr/CaRuO$_3$ is complicated by a high density of low-energy interband transitions \cite{Dang-2015}. 

Summarizing, our results demonstrate a rather remarkable consistency between experimental observations based on optical spectroscopy, resistivity and specific heat, all being consistent with $\nu=1$ Planckian behavior and $\omega/T$ scaling. We have explained the long-standing puzzle of an apparent power law of the optical spectrum over an intermediate frequency range and related the non-universal apparent exponent to the inelastic coupling constant. Looking forward, it would be valuable to extend our measurements and analysis to other cuprate compounds at doping levels close to the pseudogap quantum critical point. Our findings provide compelling evidence for the quantum critical behavior of electrons in cuprate superconductors. This raises the fundamental question of what is the nature of the associated quantum critical point, and its relation to the enigmatic pseudogap phase.

\section*{Methods}

\subsection*{Sample synthesis}
The La$_{1.76}$Sr$_{0.24}$CuO$_4$ ($p=0.24$) single crystal used in the present study was grown by the travelling solvent floating zone method~\cite{Frachet-2020}. This sample was annealed, cut and oriented along the a-b plane and polished before measuring infrared reflectivity and resistivity. 

\subsection*{Infrared optical conductivity}
We measured the infrared reflectivity from 2.5~meV to 0.5~eV using a Fourier-transform spectrometer with a home-built UHV optical flow cryostat and in-situ gold evaporation for calibrating the signal. In the energy range from 0.5 to 5~eV, we measured the real and imaginary parts of the dielectric function $\epsilon(\omega)$ using a home-built UHV cryostat installed in a visible-UV ellipsometer. Raw data for $\epsilon(\omega)$ are presented in \EDF~\ref{fig:fig6}. Combining the ellipsometry and reflectivity data and using the Kramers--Kronig relations between the reflectivity amplitude and phase, we obtained for each measured temperature the complex dielectric function in the range from 2.5~meV to 5~eV (see \SIS~\ref{app:scaling} and \EDF~\ref{fig:fig6}). The complex optical conductivity $\sigma(\omega)$ of low-energy transitions is directly linked to $\epsilon(\omega)$ by
	\begin{equation}\label{eq:sigma-from-eps}
		\sigma(\omega)=i\epsilon_0\omega\left[\epsilon_{\infty}-\epsilon(\omega)\right].
	\end{equation}
In this expression, $\epsilon_{\infty}$ is the background relative permittivity due to high-energy transitions [see Eq.~(\ref{eq:dielectric_function})]. We use international SI units, where $\epsilon_0=8.85\times10^{-5}$~kS/(cm\,THz). In the Gaussian CGS system, $\epsilon_0=1/(4\pi)$. In Sec.~\ref{sec:scaling} we propose and discuss in details a procedure to estimate the value of $\epsilon_{\infty}$. Using the value $\epsilon_{\infty}=2.76$ determined there, we display in Figs.~\ref{fig:fig1}\textbf{a} and \ref{fig:fig1}\textbf{b} the real and imaginary parts of the optical conductivity. In Fig.~\ref{fig:fig1}\textbf{a}, one observes a Drude-like behavior upon cooling from 300~K, characterized by a sharpening of the Drude peak in $\mathrm{Re}\,\sigma$ and a maximum in $\mathrm{Im}\,\sigma$ at a frequency that decreases with decreasing $T$. For temperatures below 75~K, the Drude peak is narrower than the minimum photon energy accessible with our spectrometer, 2.5~meV, which gives the impression of a gap opening in $\mathrm{Re}\,\sigma$. Yet, the superconducting transition only occurs at $T_c=19$~K. The conductivity decreases monotonically between 0.1 and 0.4~eV, before interband transitions gradually set in.

As is common for materials with strong electronic correlations, and well documented for cuprates in particular \cite{Gotze-1972, Basov-2011}, the optical conductivity has a richer frequency dependence than that of a simple Drude model. It is convenient however to consider a generalized Drude parametrization in terms of a frequency-dependent scattering rate $1/\tau(\omega)$ and mass enhancement $m^*(\omega)/m$ introduced in Eqs.~(\ref{eq:free_carrier}) and~(\ref{eq:memory}):
	\begin{equation}\label{eq:tau-and-m*}
		\sigma(\omega)=\frac{e^2K/(\hbar^2d_c)}
		{1/{\tau}(\omega)-i\omega \, m^*(\omega)/m},
	\end{equation}
so that the scattering rate and mass enhancement can be determined from the optical conductivity according to: 
	\begin{align}
		\label{eq:tau}
		\frac{1}{\tau(\omega)}&=\frac{e^2K}{\hbar^2d_c}\,\mathrm{Re}\,
		\left[\frac{1}{\sigma(\omega)}\right]\\
		\label{eq:m*}
		\frac{m^*(\omega)}{m}&=-\frac{e^2K}{\hbar^2d_c}\,\mathrm{Im}\,
		\left[\frac{1}{\omega\,\sigma(\omega)}\right].
	\end{align}
In these expressions, $d_c = 6.605$~\AA{} is the distance between two CuO$_2$ planes, $m$ is the band mass and $K$ is the spectral weight for a single plane. The determination of $m$ and $K$ is also discussed in Sec.~\ref{sec:scaling} along with that of $\epsilon_{\infty}$. $K$ only affects the absolute magnitude of $1/\tau$ and $m^*/m$, while the choice of $\epsilon_{\infty}$ has a more significant influence. 

\subsection*{DC transport experiment}
DC resistivity was measured inside a Physical Property Measurement System (PPMS) from Quantum Design in four-point geometry on the temperature range from 300~K to 2~K. The electric contacts were made by using silver wires of 50~$\mu$m and silver paste. To increase the contact quality, contacts were annealed at 500~$^{\circ}$C in oxygen atmosphere for an hour in order to get a resistance of a few ohms. To obtain the resistivity $\rho(T)$ as a function of temperature in the units $\Omega$\,cm from the raw sample resistance $R(T)$ in $\Omega$, the length $L$, width $W$, and thickness $t$ of the sample were measured to get a geometric factor $\alpha = W \times t/L$ knowing the relation: $\rho(T) = \alpha R(T)$. Resistivity was measured at two magnetic fields $H=0$~T and $H=16$~T to extract the superconducting transition temperature $T_c =19$~K at $H=0$~T and the normal-state resistivity down to 5~K ($H=16$~T).

\subsection*{Data availability}
The experimental and theoretical data generated in this study as well as the associated codes have been deposited in the Yareta database \cite{Michon-2023-data}.

\begin{acknowledgments}
A.G. acknowledges useful discussions with Andrew J. Millis, Jernej Mravlje and Subir Sachdev. We acknowledge support from the Swiss National Science Foundation under Division II through project No. 179157 (D.v.d.M.) and support through the JSPS KAKENHI grant 20H05304 (S.O.). The Flatiron Institute is a division of the Simons Foundation. L.T. acknowledges support from the Canadian Institute for Advanced Research (CIFAR) as a CIFAR Fellow and funding from the Institut Quantique, the Natural Sciences and Engineering Research Council of Canada (PIN:123817), the Fonds de Recherche du Qu\'{e}bec-Nature et Technologies (FRQNT), the Canada Foundation for Innovation (CFI), and a Canada Research Chair.
\end{acknowledgments}

\renewcommand{\thesection}{\Alph{section}}
\renewcommand{\thesubsection}{\arabic{subsection}}
\renewcommand{\theequation}{S\arabic{equation}}
\def\figurename{{Supplementary Fig.}}
\setcounter{section}{0}

\onecolumngrid
\newpage

\begin{center}\textbf{\Huge Supplementary Information\vspace{1em}}

{\large\textbf{
Reconciling scaling of the optical conductivity of cuprate superconductors\\[0.2em]with Planckian resistivity and specific heat
}}\\[1.5em]

B. Michon,$^{1, 2, 3}$ C. Berthod,$^1$ C. W. Rischau,$^1$ A. Ataei,$^4$ L. Chen,$^4$ S. Komiya,$^5$\\[0.2em]
S. Ono,$^5$ L. Taillefer,$^{4, 6}$ D. van der Marel,$^1$ and A. Georges$^{7, 8, 1, 9}$\\[0.5em]

\textit{\footnotesize
$^1$Department of Quantum Matter Physics, University of Geneva, 24 quai Ernest-Ansermet, 1211 Geneva, Switzerland\\
$^2$Department of Physics, City University of Hong Kong, 83 Tat Chee Avenue, Kowloon, Hong Kong, China\\
$^3$Hong Kong Institute for Advanced Study, City University of Hong Kong, 83 Tat Chee Avenue, Kowloon, Hong Kong, China\\
$^4$Institut Quantique, D\'{e}partement de Physique \& RQMP,\\Universit\'{e} de Sherbrooke, Sherbrooke, Qu\'{e}bec, Canada\\
$^5$Energy Transformation Research Laboratory, Central Research Institute of Electric Power Industry, 2-6-1 Nagasaka, Yokosuka, Kanagawa, Japan\\
$^6$Canadian Institute for Advanced Research, Toronto, Ontario, Canada\\
$^7$Coll\`{e}ge de France, 11 place Marcelin Berthelot, 75005 Paris, France\\
$^8$Center for Computational Quantum Physics, Flatiron Institute, New York, New York 10010, USA\\
$^9$ CPHT, CNRS, \'{E}cole Polytechnique, IP Paris, F-91128 Palaiseau, France
}

\vspace{2em}
\end{center}

\twocolumngrid

\tableofcontents

\section{Higher energy transitions and the value of $\bm{\epsilon_{\infty}}$}
\label{app:epsinv}
\let\addcontentsline\oldaddcontentsline%
\addcontentsline{toc}{section}{A.~~Higher energy transitions and the value of $\epsilon_{\infty}$}

The theoretical scaling \textit{Ansatz} that we have used to interpret the optical conductivity data only applies at low energy. Obviously, there are also higher-energy transitions that are not described by this \textit{Ansatz}. In this section, we show that these higher-energy transitions yield a contribution to $\epsilon_{\infty}$ which is of order unity. This observation is helpful in clarifying why the value of $\epsilon_{\infty}$ that provides the best possible scaling of the data is smaller than both the one deduced by analyzing the integrated spectral weight and the value more commonly admitted for this class of materials. We use a simple model of these high-energy transitions involving upper and lower Hubbard bands.

We consider electrons characterized by a spectral function of the form $A(\vec{k},\varepsilon)=A_Z(\xi_{\vec{k}},\varepsilon)+A_{\Delta}(\varepsilon)$. The term $A_Z(\xi_{\vec{k}},\varepsilon)$ represents low-energy quasiparticles with a band dispersion $\xi_{\vec{k}}$ (measured from the chemical potential) and a momentum-independent self-energy. The sum rule for this term is $\int_{-\infty}^{\infty}d\varepsilon\,A_Z(\xi_{\vec{k}},\varepsilon)=Z<1$, where $Z$ is the quasiparticle residue. The remaining spectral weight $1-Z$ is assumed to reside in lower and upper Hubbard-like bands, the former being fully occupied around energy $-\Delta_1$ and the latter being empty around energy $+\Delta_2$. For simplicity, we describes these bands as dispersion-less and write the corresponding spectral function as $A_{\Delta}(\varepsilon)=p_1\delta(\varepsilon+\Delta_1)+p_2\delta(\varepsilon-\Delta_2)$ with $\Delta_1$ and $\Delta_2$ positive. In a one-band model, the condition $p_1+p_2=1-Z$ must be obeyed to ensure the total sum rule $\int_{-\infty}^{\infty}d\varepsilon\,A(\xi_{\vec{k}},\varepsilon)=1$. In the cuprates, the hybridization of O $2p$ and Cu $3d$ electrons implies that the total spectral weight of the Hubbard bands is generally larger than $1-Z$, but difficult to estimate precisely. We will therefore keep $p_1$ and $p_2$ as independent parameters in the following.

The optical conductivity is split into a low-frequency contribution $\sigma_{\mathrm{L}}(\omega)$ determined by the transitions within the quasiparticle band and a high-frequency contribution $\sigma_{\mathrm{H}}(\omega)$ containing the transitions between the Hubbard bands and the quasiparticle band, as well as the transitions between the Hubbard bands. The dielectric function is $\epsilon(\omega)=1-\sigma(\omega)/(i\epsilon_0\omega)$. The $1$ represents the dielectric response of the vacuum with vanishing conductivity. The quantity $\epsilon_{\mathrm{L}}(\omega)=1-\sigma_{\mathrm{L}}(\omega)/(i\epsilon_0\omega)$ therefore represents the dielectric function of a vacuum dressed by the low-energy quasiparticles, without the polarizability associated with the high-frequency transitions, which is captured by $\epsilon_{\mathrm{H}}(\omega)=-\sigma_{\mathrm{H}}(\omega)/(i\epsilon_0\omega)$. If the low- and high-frequency transitions are well separated---we will assume they are---the function $\epsilon_{\mathrm{H}}(\omega)$ approaches a real constant in the low-frequency region, where the intra-band quasiparticle transitions take place, and this term can be replaced by its zero-frequency value, e.g., $\epsilon_{\mathrm{H}}(\omega)\approx\epsilon_{\mathrm{H}}(0)=\Delta\epsilon_{\infty}$ with $\infty$ recalling  that this contribution originates from high-frequency transitions. The corresponding approximate dielectric function is commonly written as $\epsilon(\omega)\approx\epsilon_{\infty}-\sigma_{\mathrm{L}}(\omega)/(i\epsilon_0\omega)$ with $\epsilon_{\infty}=1+\Delta\epsilon_{\infty}$. Our goal is to calculate $\Delta\epsilon_{\infty}$, which is formally defined as
	\begin{multline}
		\Delta\epsilon_{\infty}=\mathrm{Re}\,\epsilon_{\mathrm{H}}(0)
		=-\frac{1}{\epsilon_0}\lim_{\omega\to0}\frac{\mathrm{Im}\,\sigma_{\mathrm{H}}(\omega)}{\omega}\\
		=-\frac{1}{\epsilon_0}\lim_{\omega\to0}\frac{d}{d\omega}\mathrm{Im}\,\sigma_{\mathrm{H}}(\omega).
	\end{multline}
Using the Kramers--Kronig relation and the fact that $\mathrm{Re}\,\sigma_{\mathrm{H}}(\omega)$ is an even function of $\omega$, this may also be expressed as
	\begin{equation}\label{eq:Delta_inf}
		\Delta\epsilon_{\infty}=\frac{2}{\pi\epsilon_0}
		\int_0^{\infty}d\omega\,\frac{\mathrm{Re}\,\sigma_{\mathrm{H}}(\omega)}{\omega^2}.
	\end{equation}
This expression only makes sense if $\sigma_{\mathrm{H}}(\omega)$ vanishes sufficiently fast at $\omega=0$ or is gapped. This condition may be considered as a necessary one for the separation into low- and high-frequency degrees of freedom to be meaningful.

The optical conductivity is related to the electron spectral function and transport function $\Phi(\xi)$ via
	\begin{multline}
		\sigma(\omega)=\frac{i}{\omega}\int_{-\infty}^{\infty}d\xi\,\Phi(\xi)
		\int_{-\infty}^{\infty}d\varepsilon_1d\varepsilon_2\,A(\xi,\varepsilon_1)A(\xi,\varepsilon_2)\\
		\times\frac{f(\varepsilon_1)-f(\varepsilon_2)}{\hbar\omega+\varepsilon_1-\varepsilon_2+i0},
	\end{multline}
which reduces to Eq.~(\ref{eq:sigma}) of the main text if $\Phi(\xi)$ is replaced by $\Phi(0)$ \cite{Berthod-2013-s}. As we are not interested here in the temperature dependence of $\Delta\epsilon_{\infty}$, we set $T=0$ and deduce
	\begin{equation}
		\mathrm{Re}\,\sigma(\omega)=
		\frac{\pi}{\omega}\int_{-\infty}^{\infty}\!\!d\xi\,\Phi(\xi)
		\int_{-\hbar\omega}^0\!\!d\varepsilon\,A(\xi,\varepsilon)A(\xi,\varepsilon+\hbar\omega).
	\end{equation}
There are three contributions to $\mathrm{Re}\,\sigma_{\mathrm{H}}(\omega)$: one describes the transitions from the lower Hubbard band to the quasiparticle band [$A(\xi,\varepsilon)=p_1\delta(\varepsilon+\Delta_1)$, $A(\xi,\varepsilon+\hbar\omega)=A_Z(\xi,\varepsilon+\hbar\omega)$]; one describes the transitions from the quasiparticle band to the upper Hubbard band [$A(\xi,\varepsilon)=A_Z(\xi,\varepsilon)$, $A(\xi,\varepsilon+\hbar\omega)=p_2\delta(\varepsilon+\hbar\omega-\Delta_2)$]; one describes the transitions from the lower to the upper Hubbard bands [$A(\xi,\varepsilon)=p_1\delta(\varepsilon+\Delta_1)$, $A(\xi,\varepsilon+\hbar\omega)=p_2\delta(\varepsilon+\hbar\omega-\Delta_2)$]. The three other processes (quasiparticle to lower Hubbard, higher Hubbard to quasiparticle, and higher Hubbard to lower Hubbard) are suppressed at $T=0$. The real part of the high-frequency conductivity is, therefore,
	\begin{multline}
		\mathrm{Re}\,\sigma_{\mathrm{H}}(\omega)=
		\frac{\pi}{\omega}\int_{-\infty}^{\infty}d\xi\,\Phi(\xi)\big[\\
        p_1\theta(\hbar\omega-\Delta_1)A_Z(\xi,-\Delta_1+\hbar\omega)\\
		+p_2\theta(\hbar\omega-\Delta_2)A_Z(\xi,\Delta_2-\hbar\omega)\\
		+p_1p_2\delta(\hbar\omega-\Delta_1-\Delta_2)\big],
	\end{multline}
where $\theta(\varepsilon)$ is the Heaviside step function. $\Delta\epsilon_{\infty}$ follows from Eq.~(\ref{eq:Delta_inf}) as
	\begin{multline}
		\Delta\epsilon_{\infty}=\frac{2\hbar^2}{\epsilon_0}\int_{-\infty}^{\infty}d\xi\,\Phi(\xi)\left[
		p_1\int_0^{\infty}d\varepsilon\,\frac{A_Z(\xi,\varepsilon)}{\left(\varepsilon+\Delta_1\right)^3}\right.\\
		\left.+p_2\int_{-\infty}^0d\varepsilon\,\frac{A_Z(\xi,\varepsilon)}{\left(\Delta_2-\varepsilon\right)^3}
		+\frac{p_1p_2}{(\Delta_1+\Delta_2)^3}\right].
	\end{multline}
Since $A_Z(\xi,\varepsilon)$ is peaked near $\xi=\varepsilon$ while $\Phi(\xi)$ is a slow function of energy, we have $\int_{-\infty}^{\infty}d\xi\,\Phi(\xi)A_Z(\xi,\varepsilon)\approx\Phi(\varepsilon)\int_{-\infty}^{\infty}d\xi\,A_Z(\xi,\varepsilon)=\Phi(\varepsilon)Z$. The remaining $\varepsilon$ integral is cut at the non-interacting bandwidth $D$, beyond which $\Phi(\varepsilon)$ vanishes. For an order-of-magnitude estimate, we take $\Phi(\varepsilon)=\Phi(0)$ for $|\varepsilon|<D$ and zero otherwise, which leads to
	\begin{multline}
		\Delta\epsilon_{\infty}\sim\frac{\hbar^2\Phi(0)}{\epsilon_0\Delta_1^2}\left[
		Zp_1\frac{D(D+2\Delta_1)}{(D+\Delta_1)^2}\right.\\
		\left.+Zp_2\frac{\Delta_1^2}{\Delta_2^2}\frac{D(D+2\Delta_2)}{(D+\Delta_2)^2}
		+4p_1p_2\frac{D}{\Delta_1(1+\Delta_2/\Delta_1)^3}\right].
	\end{multline}
Using our model parameter $\Phi(0)=1.33\times10^7\epsilon_0\mathrm{THz}^2$, $\Delta_1=1.46$~eV, which corresponds to the peak in Fig.~\ref{fig:fig1}\textbf{a} of the main text, $D=4t=1.2$~eV, and $\Delta_2\gg\Delta_1, D$, we arrive at $\Delta\epsilon_{\infty}\sim 1.89Zp_1$. Hence, we see that the transitions involving the Hubbard bands yield a contribution to $\epsilon_{\infty}$ that is of order unity. A more quantitative assessment is beyond the applicability of this simple model and would also require a quantitative determination of the weights $Z$ and $p_1$.

\section{Frequency-temperature scaling of the optical data}
\label{app:scaling}

Here, we show that the frequency and temperature dependencies of the optical scattering rate and optical mass enhancement indicate a Planckian dissipation (with single-particle self-energy exponent $\nu=1$)~---~despite the fact that the modulus of the optical conductivity displays an exponent $\nu^*<1$ as seen in Fig.~\ref{fig:fig3} of the main text. Based on the local models considered in the present work and discussed in detail in \SIS~\ref{app:nu=1} and \ref{app:nu<1}, we expect that the scattering rate and mass enhancement scale according to $1/\tau(\omega)\approx T^{\nu}f_{\tau}(\omega/T)$ and $m^*(\omega)-m^*(0)\approx T^{\nu-1}f_m(\omega/T)$. If not for the approximate signs and the shift by $m^*(0)$, these scaling laws would imply that the conductivity behaves ideally as $1/\sigma(\omega)=T^{\nu}F(\omega/T)$ with $F(x)=f_{\tau}(x)-ixf_m(x)$ [see Eq.~(\ref{eq:tau-and-m*}) of the main text]. The numerical simulations show that this property is not obeyed by Eqs.~(\ref{eq:sigma}), (\ref{eq:Sigma}), and (\ref{eq:Sigma-nu}), while the approximate scaling laws with the mass shift by $m^*(0)$ are numerically well obeyed (see Fig.~\ref{fig:fig5} of the main text and \EDF~\ref{fig:fig14}). The subtraction of $m^*(0)$ in order to observe $\omega/T$ scaling is mandatory for $\nu=1$, because $m^*(0)$ varies as $\ln T$ in that case, while for $\nu<1$ it is optional, because $m^*(0)\sim T^{\nu-1}$.

\begin{figure}[tb]
\includegraphics[width=0.5\textwidth]{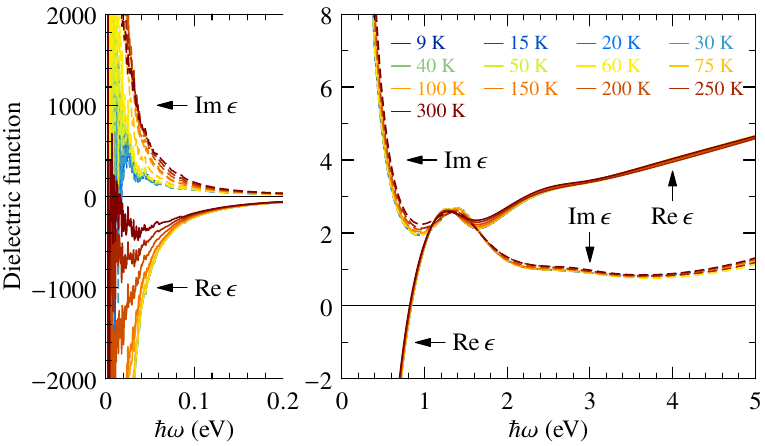}
\caption{\label{fig:fig6}
\textbf{\boldmath Complex dielectric function $\epsilon$ of LSCO at $p=0.24$.}
These data are obtained from a mix of two techniques, namely the infrared reflectivity between 2.5~meV and 0.5~eV and ellipsometry from 0.5 to 5~eV. The part 0.5--5~eV is using a fit to the ellipsometry in order to extrapolate the reflectivity data at higher photon energy and extract $\epsilon$ through Kramers--Kronig relations. 
}
\end{figure}

\begin{figure*}[t]
\includegraphics[width=0.9\textwidth]{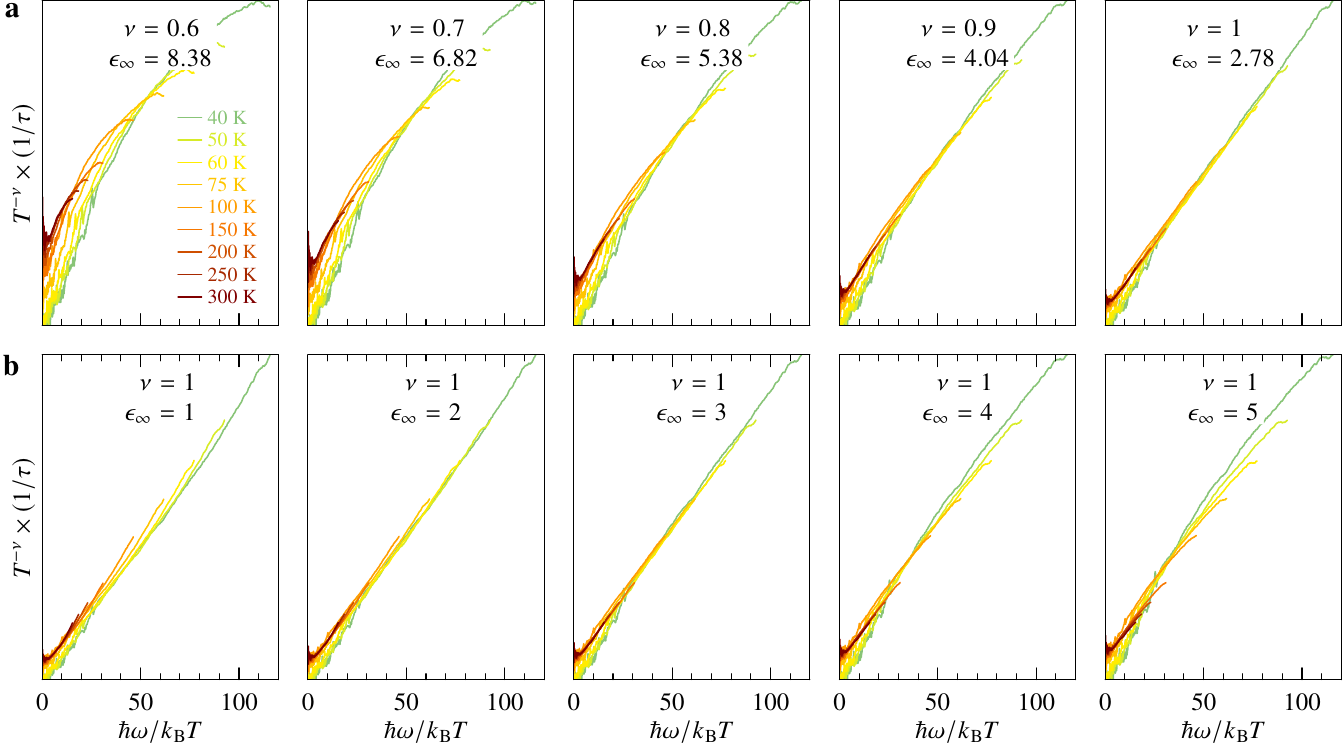}
\caption{\label{fig:fig7}
\textbf{Evidence of Planckian scaling for the scattering rate.}
\textbf{a} Optimal $\omega/T$ scaling of the scattering rate for the values of $\nu$ indicated in each panel; the value of $\epsilon_{\infty}$ is the one ensuring the best collapse of the curves. \textbf{b} Sensitivity of the scaling collapse to the value of $\epsilon_{\infty}$ for $\nu=1$. The vertical axis starts at zero in each panel and is otherwise arbitrary, unless a value of the spectral weight $K$ is specified.
}
\end{figure*}

The measured infrared dielectric function of LSCO at doping $p = 0.24$ is displayed in \EDF~\ref{fig:fig6}. We emphasize in the main text that the frequency dependence of $1/\tau$ and $m^*/m$ extracted from the dielectric function depends on the value chosen for the background dielectric constant. Here, in order to test the presence of scaling laws in the data, we optimize the value of $\epsilon_{\infty}$ to achieve the best scaling collapse. The results of this procedure for $1/\tau$ are displayed in \EDF~\ref{fig:fig7}\textbf{a}. For each value of $\nu$, we minimize a cost function representing the quality of the collapse and we deduce the optimal $\epsilon_{\infty}$ indicated in each panel. We use the optical data in the range $T>30$~K and $\hbar\omega<0.4~\mathrm{eV}$ for this analysis. The data for $T=30$~K and below are strongly affected by the loss of information due to the sharpening of the Drude peak below our observation limit of 2.5~meV and the opening of the superconducting gap below $T_c=19$~K. Furthermore, interband transitions come into play above $0.4~\mathrm{eV}$. The figure shows that the collapse improves upon increasing $\nu$ towards $\nu=1$. For $\nu\lesssim0.8$, the optimal $\epsilon_{\infty}$ is larger than the largest measured infrared dielectric function (\EDF~\ref{fig:fig6}) and the collapse is poor. \EDF~\ref{fig:fig7}\textbf{b} illustrates how the scaling collapse for $\nu=1$ depends on the value of $\epsilon_{\infty}$.

\begin{figure*}[t]
\includegraphics[width=0.9\textwidth]{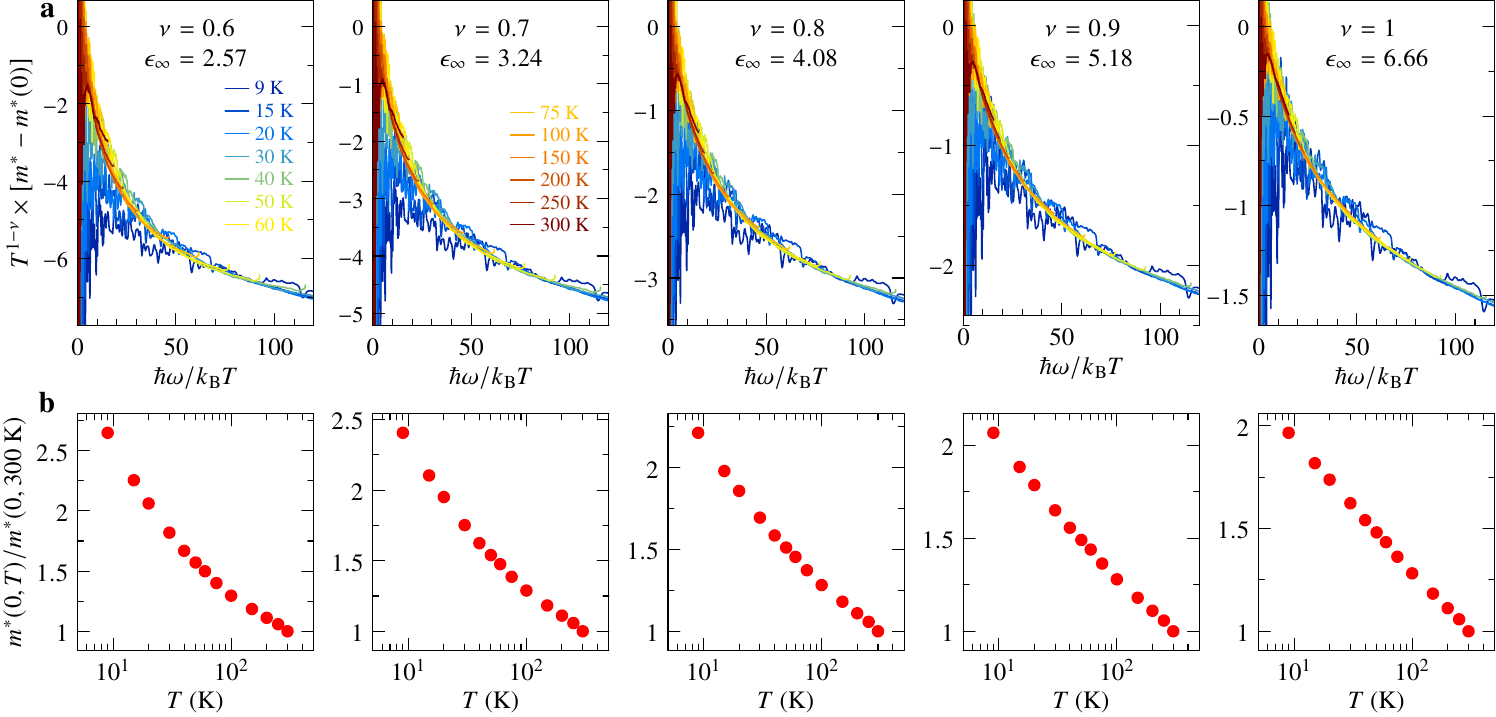}
\caption{\label{fig:fig8}
\textbf{Evidence of Planckian scaling for the mass enhancement.}
\textbf{a} Optimal $\omega/T$ scaling of the mass enhancement for the values of $\nu$ indicated in each panel; the value of $\epsilon_{\infty}$ is the one ensuring the best collapse of the curves. \textbf{b} Values of $m^*(0)$ that give the best collapse.
}
\end{figure*}

In order to test the scaling laws for $m^*$, we must determine $m^*(0)$. The low-frequency noise prevents us from extracting this information directly from the data. We therefore treat $m^*(0)$ for each temperature as a variable that we adjust, like $\epsilon_{\infty}$, for optimal collapse. The results are displayed in \EDF~\ref{fig:fig8}. \EDF~\ref{fig:fig8}\textbf{a} shows a systematic improvement of the collapse with $\nu$ approaching unity. All temperatures were included in this analysis, which allows us to get the full $T$-dependence of $m^*(0)$, as shown in \EDF~\ref{fig:fig8}\textbf{b}. Note that this procedure does not set an absolute scale for $m^*(0)$, unless the scale of $m^*$ is fixed by a choice of the spectral weight $K$. For $\nu=1$, the optimal $m^*(0)$ values align on a straight line, indicating a precise $\ln T$ behavior consistent with the behavior of the quasiparticle mass revealed by the electronic specific heat.

The optimizations presented so far establish that the self-energy exponent takes the value $\nu=1$. However, the optimal $\epsilon_{\infty}$ determined from $1/\tau$ and $m^*$ differ. One sees in \EDF~\ref{fig:fig7}\textbf{b} that the collapse of $1/\tau$ quickly deteriorates as $\epsilon_{\infty}$ increases beyond 3 and would be badly broken for $\epsilon_{\infty}=6.66$, as required for optimal collapse of the mass. In fact, the large values of $\epsilon_{\infty}$ stem from the $m^*$ data above $\sim 0.2$~eV. It is seen in Fig.~\ref{fig:fig1}\textbf{d} of the main text that a change of behavior occurs around 0.22~eV in the mass enhancement data. The data above 0.22~eV scale in a different way, as also seen in Fig.~\ref{fig:fig2}\textbf{d} of the main text. We find that the optimal $\epsilon_{\infty}$ drops to a value close to 3 if those data are ignored. In contrast, the optimal $\epsilon_{\infty}$ for $1/\tau$ does not change appreciably upon varying the data range below 0.4~eV. We therefore perform a common optimization for both $1/\tau$ and $m^*$ by minimizing the sum of the two cost functions (normalized to the value at their respective minimum), considering energies up to 0.4~eV and temperatures larger than 30~K for $1/\tau$, while keeping all temperatures but only energies up to 0.22~eV for $m^*$. The resulting optimum is $\epsilon_{\infty}=2.76$, very close to the optimum for $1/\tau$, and the resulting zero-frequency masses and scaling collapses are displayed in Fig.~\ref{fig:fig2} of the main text. Note that these technical choices are inessential: by keeping data up to 0.22~eV or 0.4~eV for both $1/\tau$ and $m^*$, we obtain very similar optimal $\epsilon_{\infty}$ of 2.91 and 3.03, respectively, and virtually identical values of $m^*(0)$.

\let\oldaddcontentsline\addcontentsline%
\renewcommand{\addcontentsline}[3]{}%
\section{Study of the Planckian model ($\bm{\nu=1}$)}
\label{app:nu=1}
\let\addcontentsline\oldaddcontentsline%
\addcontentsline{toc}{section}{C.~~Study of the Planckian model ($\nu=1$)}

Here, we study the Planckian model defined by Eqs.~(\ref{eq:Sigma}) and (\ref{eq:sigma}) of the main text. We calculate the quasiparticle mass, derive approximate analytical expressions for the self-energy and the frequency-dependent optical conductivity, and check these approximations against the numerically computed conductivity. We finally discuss the $\omega/T$ scaling properties of the model.

\subsection{Single-particle self-energy}

We implement the ultraviolet cutoff in Eq.~(\ref{eq:Sigma}) by limiting the imaginary part of the self-energy to the constant value $\mathrm{Im}\,\Sigma(\Lambda)=-g\pi \kB TS(\Lambda/\kB T)$ for all energies $|\varepsilon|>\Lambda$. The real part contains a constant term $\mathrm{Re}\,\Sigma(0)$ that is compensated by a shift of chemical potential and can therefore be subtracted from $\Sigma(\varepsilon)$. The remaining real part describes the renormalization of the single-particle dispersion by inelastic scattering processes, which at leading order amounts to a renormalization of the mass. The renormalized single-particle mass or quasiparticle mass is 
	\begin{equation}
		\frac{m^*_{\mathrm{qp}}}{m}
		=1-\frac{d}{d\varepsilon}\mathrm{Re}\left[\Sigma(\varepsilon)-\Sigma(0)\right]_{\varepsilon=0}
		\equiv\frac{1}{Z},
	\end{equation}
where $Z$ is the quasiparticle residue. After evaluating the derivative of $\Sigma(\varepsilon)$ using Eq.~(\ref{eq:Sigma}), performing an integration by parts, and separating the even and odd terms in the resulting integral, we arrive at the expression:
	\begin{align}
		\nonumber
		\frac{m^*_{\mathrm{qp}}}{m}&=1+g\int_0^{\frac{\Lambda}{\kB T}}dx\,\frac{S'(x)-S'(-x)}{x}\\
		&=1+2g\int_0^{\frac{\Lambda}{\kB T}}dx\,\frac{\sinh x-x}{x(\cosh x-1)}.
	\end{align}
The function to integrate approaches $1/x$ at large $x$ and the integral therefore approaches $\ln(\Lambda/\kB T)+\mathrm{cste}$ if $\Lambda\gg \kB T$. We write this as
	\begin{equation}\label{eq:mqp}
		\frac{m^*_{\mathrm{qp}}}{m}=1+2g\ln\left(a\frac{\Lambda}{\kB T}\right)
		\qquad(\Lambda\gg \kB T),
	\end{equation}
where the constant $a$ is given by
	\begin{equation*}
		a=\lim_{x_c\to\infty}\frac{1}{x_c}\exp\left[\int_0^{x_c}dx\,\frac{\sinh x-x}{x(\cosh x-1)}\right]
		=0.770542.
	\end{equation*}
If $\varepsilon\gg \kB T$, the integral giving $\Sigma(\varepsilon)$ is dominated by the region $|x|\gg1$ and we can replace the scaling function by its asymptotic form $S(|x|\gg1)=|x|$. We then find
	\begin{equation}\label{eq:ReSigma1}
		\mathrm{Re}\left[\Sigma(\varepsilon)-\Sigma(0)\right]
		\approx g\varepsilon\ln\left|\frac{\varepsilon^2}{\Lambda^2-\varepsilon^2}\right|
		+g\Lambda\ln\left|\frac{\Lambda-\varepsilon}{\Lambda+\varepsilon}\right|
	\end{equation}
in this regime, which will be used below in building approximations for the conductivity.

\subsection{Optical conductivity}

\begin{figure}[tb]
\includegraphics[width=0.5\textwidth]{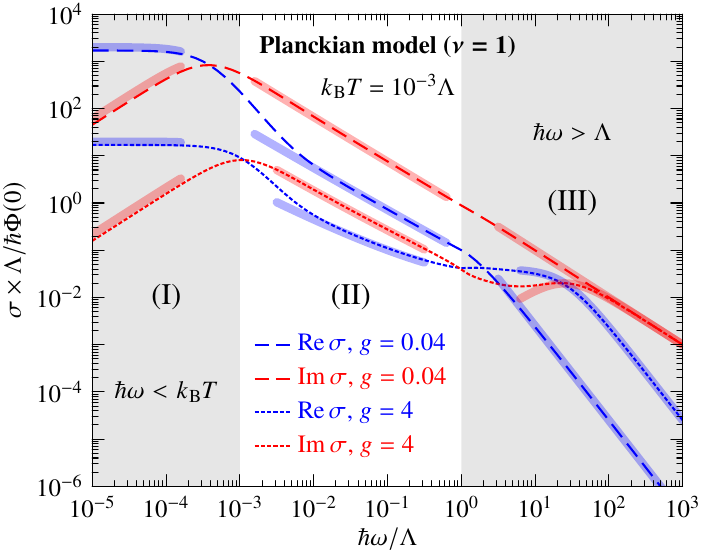}
\caption{\label{fig:fig9}
\textbf{Optical conductivity of a Planckian model.}
Real (blue) and imaginary (red) parts of the optical conductivity, Eqs.~(\ref{eq:sigma}) and (\ref{eq:Sigma}) of the main text, for $g=0.04$ (dashed) and $g=4$ (dotted) at temperature $\kB T=10^{-3}\Lambda$. The thick lines show Eqs.~(\ref{eq:sigma1}), (\ref{eq:sigma2}), and (\ref{eq:sigma3}) in the corresponding regime of frequency.
}
\end{figure}

\EDF~\ref{fig:fig9} shows the optical conductivity calculated for weak coupling ($g=0.04$) and strong coupling ($g=4$) at a temperature $\kB T=10^{-3}\Lambda$, and plotted using the cutoff $\Lambda$ as the unit of energy. One can distinguish three regimes of frequency, as pointed out in Ref.~\onlinecite{vanderMarel-2003-s}.

In the regime (I), the frequency is lower than the temperature. The scattering rate approaches a frequency-independent value proportional to $T$ and the conductivity approaches a Drude form with renormalized spectral weight $Z\Phi(0)$. To obtain the asymptotic expression of the conductivity, we expand numerator and denominator in Eq.~(\ref{eq:sigma}) around $\omega=0$ to get
	\begin{equation*}
		\sigma_{\mathrm{(I)}}(\omega)=i\hbar\Phi(0)\int_{-\infty}^{\infty}d\varepsilon\,
		\frac{-df(\varepsilon)/d\varepsilon}{\hbar\omega\left[1-d\Sigma(\varepsilon)/d\varepsilon\right]
		-2i\mathrm{Im}\,\Sigma(\varepsilon)}.
	\end{equation*}
At low enough temperature, the derivative of the Fermi function is sharply peaked at $\varepsilon=0$ and the denominator in the function to integrate can be approximated by its value at $\varepsilon=0$. Since the imaginary part of the self-energy is an even function of $\varepsilon$, the term in square brackets becomes $1-d\mathrm{Re}\,\Sigma(\varepsilon)/d\varepsilon|_{\varepsilon=0}=1/Z$. Furthermore, Eq.~(\ref{eq:Sigma}) gives $\mathrm{Im}\,\Sigma(0)=-2\pi g \kB T$, leading to
	\begin{equation}\label{eq:sigma1}
		\sigma_{\mathrm{(I)}}(\omega)\approx\frac{Z\Phi(0)}{-i\omega+4\pi g Z \kB T/\hbar}.
	\end{equation}
This has the usual dissipative ($\mathrm{Re}\,\sigma>\mathrm{Im}\,\sigma$) Drude structure with a scattering time $\tau=\hbar/(4\pi g \kB T)$, corresponding to a $T$-linear resistivity. The coefficient $A$ of the resistivity deduced from Eq.~(\ref{eq:sigma1}) is smaller than the exact result, Eq.~(\ref{eq:A}) of the main text, by a factor $7\zeta(3)/\pi^2=0.853$. The predictions of Eq.~(\ref{eq:sigma1}) with $1/Z$ given by Eq.~(\ref{eq:mqp}) are displayed in \EDF~\ref{fig:fig9} as the thick lines in the regime (I).

\begin{figure}[b]
\includegraphics[width=0.4\textwidth]{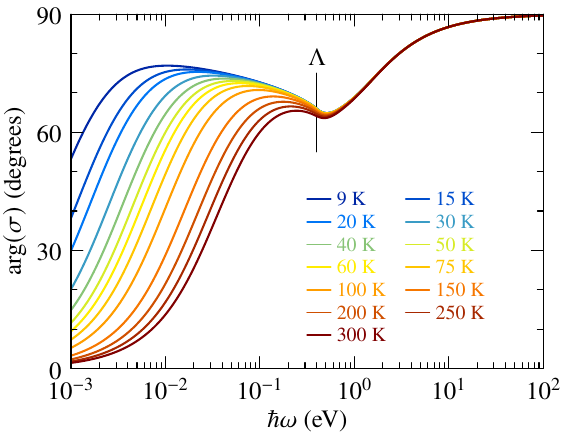}
\caption{\label{fig:fig10}
\textbf{\boldmath Full frequency dependence of $\arg(\sigma)$.}
Crossover from the regime of effective exponent $\nu^*<1$ for $\hbar\omega<\Lambda$ to the asymptotic regime showing $\nu=1$ and $\arg(\sigma)=\pi/2$ for $\hbar\omega>\Lambda$. The model parameters are $g=0.23$ and $\Lambda=0.4$~eV.
}
\end{figure}

In the regimes (II) and (III), the frequency is large compared with temperature and we can therefore replace the Fermi functions by their expression at $T=0$:
	\begin{align}\label{eq:sigma23}
		\nonumber
		\sigma(\hbar\omega\gg \kB T)&=\frac{i\Phi(0)}{\omega}\int_{-\hbar\omega}^0
		\frac{d\varepsilon}{\hbar\omega+\Sigma^*(\varepsilon)-\Sigma(\varepsilon+\hbar\omega)}\\
		&\approx\frac{i\hbar\Phi(0)}{\hbar\omega-2\Sigma(\hbar\omega/2)}.
	\end{align}
At the second line, we have approximated the function to integrate by its value in the middle of the integration window, noting that the real (imaginary) part of $\Sigma(\varepsilon)$ is odd (even) in $\varepsilon$~---~provided that the constant $\mathrm{Re}\,\Sigma(0)$ is subtracted from $\Sigma(\varepsilon)$, which is implicit in Eq.~(\ref{eq:sigma23}). The condition $\hbar\omega\gg \kB T$ allows us to use Eq.~(\ref{eq:ReSigma1}) and the asymptotic form $S(x)=|x|$ when evaluating $\mathrm{Im}\,[-2\Sigma(\hbar\omega/2)]=\pi g\min(\hbar\omega,2\Lambda)$. In the regime (III), the real part of $\Sigma(\hbar\omega/2)$ disappears from the conductivity because the right-hand side of Eq.~(\ref{eq:ReSigma1}) drops as $\Lambda/\varepsilon$, such that we arrive at the following approximation:
	\begin{equation}\label{eq:sigma3}
		\sigma_{\mathrm{(III)}}(\omega)\approx\frac{\Phi(0)}{-i\omega
		+\pi g\min\left(\omega,2\Lambda/\hbar\right)}.
	\end{equation}
This is asymptotically an inductive regime ($\mathrm{Im}\,\sigma>\mathrm{Re}\,\sigma$), where the real part decreases as $1/\omega^2$ and the imaginary part as $1/\omega$, such that $|\sigma_{\mathrm{(III)}}|\sim1/\omega$. As the asymptotic conductivity is purely imaginary, $\mathrm{arg}(\sigma_{\mathrm{(III)}})$ approaches the value $\pi/2$. This is illustrated in \EDF~\ref{fig:fig10}. Equation~(\ref{eq:sigma3}) agrees with the numerics, as the thick lines show in the regime (III) of \EDF~\ref{fig:fig9}.

For the regime (II), we expand Eq.~(\ref{eq:ReSigma1}) for $\hbar\omega\ll\Lambda$ and arrive at the expression
	\begin{equation}\label{eq:sigma2}
		\sigma_{\mathrm{(II)}}(\omega)\approx\frac{\Phi(0)}{-i\omega}\frac{1}{1
		+2g\left[1-\ln\left(\frac{\hbar\omega}{2\Lambda}\right)\right]
		+i\pi g},
	\end{equation}
which is also in good agreement with the numerics (thick lines in the regime (II) of \EDF~\ref{fig:fig9}). Without the cutoff-dependent logarithmic correction, $\sigma_{\mathrm{(II)}}(\omega)$ would display pure $\sigma\propto(-i\omega)^{-1}$ behavior, as may be expected in a quantum critical system with linear-in-energy single-particle scattering rate \cite{vanderMarel-2003-s}. This is realized for $g\to0$, as shown by the dashed lines in \EDF~\ref{fig:fig9}. For finite $g$, however, the logarithmic correction reduces the decay rate of both real and imaginary parts (dotted lines).

We see in \EDF~\ref{fig:fig9} that the approximation Eq.~(\ref{eq:sigma2}) captures the change of behavior of the conductivity with increasing $g$ in the intermediate frequency regime $\kB T<\hbar\omega<\Lambda$. We therefore use this approximation in order to extract the apparent power-law exponent of $|\sigma|$ by computing the logarithmic derivative at $\hbar\omega=\Lambda/2$, near the middle of domain (II). We thus arrive at Eq.~(\ref{eq:nustar}) of the main text.

\let\oldaddcontentsline\addcontentsline%
\renewcommand{\addcontentsline}[3]{}%
\subsection{\bm{$\omega/T$} scaling}
\let\addcontentsline\oldaddcontentsline%
\addcontentsline{toc}{subsection}{3.~~$\omega/T$ scaling}

As the approximations presented so far take one of the two limits $\omega\ll T$ or $\omega\gg T$, they cannot predict the form of the $\omega/T$ scaling expected when $\omega$ and $T$ are comparable. If the self-energy were obeying the scaling property $\Sigma(\varepsilon)=T\mathscr{F}(\varepsilon/T)$, then also the conductivity, Eq.~(\ref{eq:sigma}), would obviously scale exactly like $1/\sigma(\omega)=TF(\omega/T)$. The cutoff that must be introduced in Eq.~(\ref{eq:Sigma}) breaks this property, since the self-energy is rather of the form $\Sigma(\varepsilon)=T\mathscr{F}(\varepsilon/T,\Lambda/T)$. The perfect scaling of $1/\sigma$ is therefore lost as well. In the following, we obtain closed formula for the approximate scaling of $1/\sigma$.

At low energy and temperature, $\mathrm{Im}\,\Sigma$ is not influenced by the cutoff, such that the ideal scaling $1/\sigma\propto-i\omega m^*/m+1/\tau=TF(\omega/T)$ is expected to hold reasonably well for the real part, i.e., $1/\tau\approx Tf_{\tau}(\omega/T)$. This expectation is confirmed by the numerical simulations (Fig.~\ref{fig:fig5}\textbf{a} of the main text). Equation~(\ref{eq:sigma1}) indicates that the function $f_{\tau}(x)$ is close to $4\pi g$ at $x=0$, which is confirmed as well by the numerics. Since this corresponds to $-2\mathrm{Im}\,\Sigma(0)/\kB T$, it is tempting to infer that $1/\tau$ is close to twice the single-particle scattering rate, which would give $2\pi gS(x)$ for the scaling function. However, that function goes to $2\pi gx$ at large $x$, while the simulations go to $\pi gx$. We therefore try $f_{\tau}(x)=2\pi gS(x/2)$, which works well despite small deviations at low $x$, as seen in Fig.~\ref{fig:fig5}\textbf{a} of the main text. This result could have been directly guessed from Eq.~(\ref{eq:sigma23}).

\begin{figure}[tb]
\includegraphics[width=0.4\textwidth]{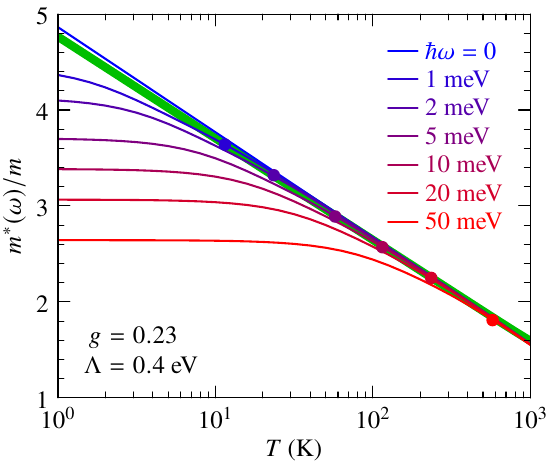}
\caption{\label{fig:fig11}
\textbf{Optical mass versus quasiparticle mass.}
Temperature-dependent optical mass enhancement predicted by Eqs.~(\ref{eq:sigma}) and (\ref{eq:Sigma}) of the main text for $g=0.23$, $\Lambda=0.4$~eV and various frequencies, compared with the quasiparticle mass from Eq.~(\ref{eq:mqp}) in green. The dots indicate $\hbar\omega=\kB T$. For $\kB T>\hbar\omega$, we have $m^*(\omega)\approx m^*_{\mathrm{qp}}$.
}
\end{figure}

Since Eq.~(\ref{eq:sigma23}) correctly predicts the approximate scaling of $1/\tau$, we use this same relation for $m^*$, which leads to $m^*/m\approx 1-(2/\hbar\omega)\mathrm{Re}[\Sigma(\hbar\omega/2)-\Sigma(0)]$. This expression immediately gives $m^*(0)/m\approx1-\Sigma'(0)=m^*_{\mathrm{qp}}/m$, a result confirmed by the numerics (see \EDF~\ref{fig:fig11}). If the function $S(x)$ is replaced by the very similar but simpler function $|x|+2e^{-|x|/2}$, which has the same behavior at low and high values of $x$, a complete evaluation of $\mathrm{Re}[\Sigma(\varepsilon)-\Sigma(0)]$ becomes possible. We thus find that $m^*(\omega)-m^*(0)$ scales as $f_m(\omega/T)$ with
	\begin{equation}\label{eq:fm}
		f_m(x)=2g\left\{1\!-\!\gamma\!-\!\ln\left(\frac{x}{4}\right)\!+\!\frac{2}{x}\left[
		e^{\frac{x}{4}}\mathrm{Ei}\left(-\frac{x}{4}\right)
		\!-\!e^{-\frac{x}{4}}\mathrm{Ei}\left(\frac{x}{4}\right)\right]\right\}.
	\end{equation}
$\gamma=0.577$ is Euler's constant and $\mathrm{Ei}$ is the exponential integral function. The function $f_m$ is compared in Fig.~\ref{fig:fig5}\textbf{b} of the main text with the numerical simulations. The slight difference is a consequence of using the approximation Eq.~(\ref{eq:sigma23}) for computing the optical mass, not a consequence of approximating the function $S(x)$.

\begin{figure}[tb]
\includegraphics[width=0.5\textwidth]{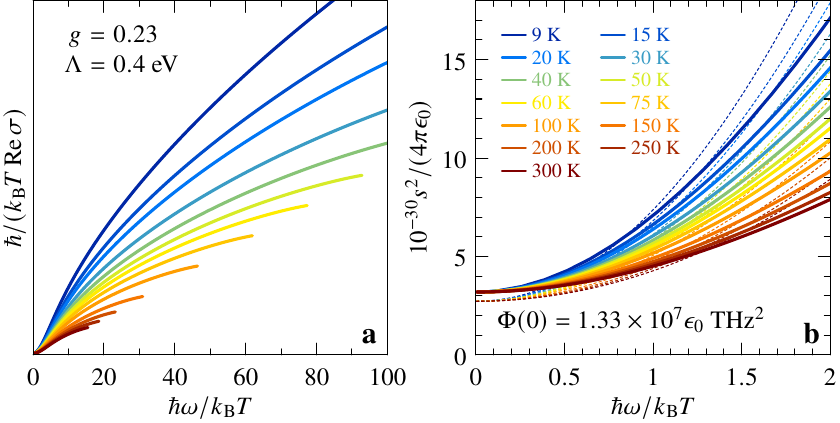}
\caption{\label{fig:fig12}
\textbf{\boldmath Scaling violation in $\mathrm{Re}\,\sigma$.}
Absence of $\omega/T$ scaling in $\hbar/(\kB T\,\mathrm{Re}\,\sigma)$ evaluated numerically in the Planckian model over \textbf{a} a wide and \textbf{b} a narrow range of $\omega/T$. The dotted lines in \textbf{b} show the approximation Eq.~(\ref{eq:kTResigma}).
}
\end{figure}

To conclude this section, we note that the presence of $\omega/T$ scaling has been previously tested in Bi2212 by considering the quantity $\hbar/(\kB T\mathrm{Re}\,\sigma)$ \cite{vanderMarel-2003-s}. Equation (\ref{eq:sigma-from-eps}) of the main text shows that $\mathrm{Re}\,\sigma$ is independent of $\epsilon_\infty$ and $K$, making it the ideal quantity for testing the presence of $\omega/T$ scaling in the raw data. In our model, however, this quantity does not scale with $\omega/T$, as seen in \EDF~\ref{fig:fig12}. Using the form $\sigma\propto1/(-i\omega\,m^*/m+1/\tau)$ and our observations that, to a very good accuracy, $\hbar/\tau=\kB Tf_{\tau}(\omega/T)$ and $m^*/m=m^*_{\mathrm{qp}}/m+f_m(\omega/T)$, we deduce
	\begin{equation*}
		\frac{\hbar}{\kB T\mathrm{Re}\,\sigma}\sim
		f_{\tau}\left(\frac{\omega}{T}\right)+\frac{(\hbar\omega/\kB T)^2}
		{f_{\tau}(\omega/T)}\left[\frac{m^*_{\mathrm{qp}}}{m}+f_m\left(\frac{\omega}{T}\right)\right]^2.
	\end{equation*}
Due to the logarithmic variation of $m^*_{\mathrm{qp}}$ with temperature, the second term on the right-hand side is not a function of $\omega/T$. Despite the conspicuous absence of scaling, the behavior of the model in a narrow range of $\omega/T$ (\EDF~\ref{fig:fig12}\textbf{b}) and for temperatures above 100~K is strikingly similar to that reported in Ref.~\onlinecite{vanderMarel-2003-s}, suggesting that the model could be useful to analyze Bi2212 data as well. In particular, $\hbar/(\kB T\,\mathrm{Re}\,\sigma)$ increases as $(\omega/T)^2$ as pointed out in Ref.~\onlinecite{vanderMarel-2003-s}, albeit with a $T$-dependent curvature. Since this curvature is in principle accessible in the raw optical data, it is interesting to relate it to the parameters of our Planckian model. Using our approximate scaling functions, we find for $\omega\to0$:
	\begin{equation}\label{eq:kTResigma}
		\frac{\hbar}{\kB T\mathrm{Re}\,\sigma}\approx\frac{4\pi g}{\Phi(0)}
		\left\{1+\left[\frac{1}{48}+\left(\frac{m^*_{\mathrm{qp}}/m}{4\pi g}\right)^2
		\right]\left(\frac{\hbar\omega}{\kB T}\right)^2\right\}.
	\end{equation}
This expression is in reasonable agreement with the exact numerical result (see dotted lines in \EDF~\ref{fig:fig12}\textbf{b}). Equation~(\ref{eq:kTResigma}) can in principle be used to extract the three parameters of the theory all at once from the raw optical data. This endeavor is somewhat risky, though, because the approximate validity of Eq.~(\ref{eq:kTResigma}) is limited to a very narrow low-frequency range, where optical data is usually noisy. Furthermore, the sensitivity to the cutoff $\Lambda$ is only logarithmic.

\let\oldaddcontentsline\addcontentsline%
\renewcommand{\addcontentsline}[3]{}%
\section{Study of the Sub-Planckian model ($\bm{\nu<1}$)}
\label{app:nu<1}
\let\addcontentsline\oldaddcontentsline%
\addcontentsline{toc}{section}{D.~~Study of the Sub-Planckian model ($\nu<1$)}

Here, we introduce and study a model with sub-linear energy and temperature dependencies of the self-energy. We follow the same outline as for the Planckian model, first discussing the self-energy, then the optical conductivity, and finally the $\omega/T$ scaling. Beside presenting the model, our main goal is to argue that it gives predictions that are in disagreement with the experimental observations in LSCO at doping $p^*$, \emph{including} for the apparent exponent $\nu^*<1$ of the optical conductivity.

\subsection{Single-particle self-energy}

The central property of the class of theories discussed in this paper is that the single-particle scattering rate is local (independent of momentum) and assumes the form $-\mathrm{Im}\,\Sigma(\varepsilon)=\pi g(\kB T)^{\nu}S_{\nu}(\varepsilon/\kB T)$, where $S_{\nu}(0)$ is finite and $S_{\nu}(|x|\gg1)\propto|x|^{\nu}$. These two conditions ensure that the dc resistivity behaves as $T^{\nu}$ and the ac scattering rate approaches $\omega^{\nu}$ for $\hbar\omega> \kB T$, while at intermediate frequencies the scattering rate divided by $T^{\nu}$ is a function of $\omega/T$. The case $\nu=1$ is the Planckian model described in \SIS~\ref{app:nu=1}, while here we consider the case $\nu<1$. An important aspect of the sub-Planckian model is that it does not require an ultraviolet cutoff, because the Kramers--Kronig integral giving the real part of the self-energy is now convergent. Thus the theory is completely specified in terms of the low-energy properties of the carriers, which then determine the thermodynamic, transport, and spectroscopic properties. This is an important difference relative to the case $\nu=1$, where the high-energy structure of the theory, represented by the ultraviolet cutoff, controls the thermodynamics and the optical mass.

Certain microscopic models with conformal invariance realize exactly the scaling form of the single-particle scattering rate with $\nu<1$ and provide an explicit expression for the function $S_{\nu}(x)$ \cite{Affleck-1991, Cox-1998, Parcollet-1998-s, Dumitrescu-2022-s}. We borrow our self-energy model from those:
	\begin{subequations}\label{eq:Sigma-nu}\begin{align}
		\Sigma(z)&=g(\kB T)^{\nu}\int_{-\infty}^{\infty}dx\,
		\frac{S_{\nu}(x)}{z/\kB T-x}\\
		S_{\nu}(x)&=\frac{(2\pi)^{\nu}}{\pi\,\Gamma(1+\nu)}\cosh(x/2)
		\left|\,\Gamma\left(\frac{1+\nu}{2}+i\frac{x}{2\pi}\right)\right|^2,
	\end{align}\end{subequations}
where $\Gamma$ denotes the Euler gamma function. Note that $g$ is dimensionfull with the unit of energy to the power $1-\nu$. Proceeding like in the case $\nu=1$, we find that the quasiparticle mass diverges at low temperature like $T^{\nu-1}$:
	\begin{equation*}
		\frac{m^*_{\mathrm{qp}}}{m}=1+g(\kB T)^{\nu-1}c_{\nu},\quad
		c_{\nu}=\int_0^{\infty}dx\,\frac{S_{\nu}'(x)-S_{\nu}'(-x)}{x}.
	\end{equation*}
This is a robust property of the sub-Planckian model~---~it does not depend on the particular scaling function $S_{\nu}(x)$~---~that disagrees with the logarithmic temperature dependence observed experimentally in LSCO at doping $p = 0.24$. Arguably, for $\nu$ close to unity it is difficult to distinguish the power law $T^{\nu-1}$ from a logarithmic temperature dependence.

In order to derive approximations for the conductivity, it is helpful to estimate the self-energy at energies larger than $\kB T$. For $\varepsilon\gg \kB T$, we replace $S_{\nu}(x)$ by its asymptotic value, which is $|x|^{\nu}/\Gamma(1+\nu)$, and we obtain for the real part:
	\begin{equation}\label{eq:ReSigma-nu}
		\mathrm{Re}[\Sigma(\varepsilon)-\Sigma(0)]=-g\frac{\pi\tan(\pi\nu/2)}{\Gamma(1+\nu)}
		\mathrm{sign}(\varepsilon)|\varepsilon|^{\nu}\quad(\varepsilon\gg \kB T).
	\end{equation}

\let\oldaddcontentsline\addcontentsline%
\renewcommand{\addcontentsline}[3]{}%
\subsection{Optical conductivity and $\bm{\omega/T}$ scaling}
\let\addcontentsline\oldaddcontentsline%
\addcontentsline{toc}{subsection}{2.~~Optical conductivity and $\omega/T$ scaling}

\begin{figure}[tb]
\includegraphics[width=0.5\textwidth]{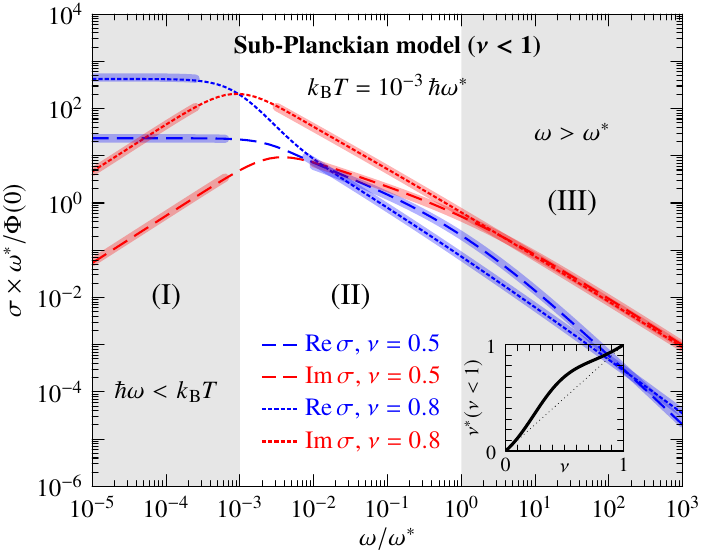}
\caption{\label{fig:fig13}
\textbf{Optical conductivity of a sub-Planckian model.}
Real (blue) and imaginary (red) parts of the optical conductivity, Eqs.~(\ref{eq:sigma}) of the main text and (\ref{eq:Sigma-nu}), for $\nu=0.5$ (dashed) and $\nu=0.8$ (dotted) at temperature $\kB T=10^{-3}\,\hbar\omega^*$. The thick lines show Eqs.~(\ref{eq:sigma1-nu}) and (\ref{eq:sigma23-nu}). Inset: effective exponent as given by Eq.~(\ref{eq:nustar-nu}).
}
\end{figure}

The frequency-dependent conductivity crosses over from the dissipative Drude regime to the asymptotic inductive regime via an intermediate regime whose extension is controlled by $g$ and $\nu$ (while for $\nu=1$, it is controlled by the cutoff $\Lambda$). We define a crossover frequency $\omega^*$ by the condition $\hbar\omega^*=2\mathrm{Re}[\Sigma(\hbar\omega^*)-\Sigma(0)]$: for $\omega<\omega^*$, the self-energy dominates in Eq.~(\ref{eq:sigma}) while for $\omega>\omega^*$, one approaches the regime where $\mathrm{Re}\,\Sigma$ becomes irrelevant. Using Eq.~(\ref{eq:ReSigma-nu}), we find
	\begin{equation}
		\hbar\omega^*=\left[\frac{2\pi g\tan(\pi\nu/2)}{\Gamma(1+\nu)}\right]^{\frac{1}{1-\nu}}.
	\end{equation}
If the temperature and the frequency are both measured in units of $\omega^*$, the conductivity multiplied by $\omega^*$ becomes a function of $\omega/\omega^*$ and $T/\omega^*$ that depends on $\nu$, but no longer on $g$. This function is displayed in \EDF~\ref{fig:fig13} for two values of $\nu$. The change of behavior around $\omega^*$ is clearly visible.

In the Drude regime, we proceed like for $\nu=1$ and get
	\begin{equation}\label{eq:sigma1-nu}
		\sigma_{\mathrm{(I)}}(\omega)\approx\frac{Z\Phi(0)}{-i\omega+2gZ
		\frac{\left[\Gamma\left(\frac{1+\nu}{2}\right)\right]^2}{\Gamma(1+\nu)}
		(2\pi \kB T)^{\nu}/\hbar},
	\end{equation}
as indicated by the thick lines in the regime (I) of \EDF~\ref{fig:fig13}. This shows that the resistivity is proportional to $T^{\nu}$, another robust property of the sub-Planckian model that disagrees with the linear resistivity observed in LSCO at $p = 0.24$.

\begin{figure}[tb]
\includegraphics[width=0.5\textwidth]{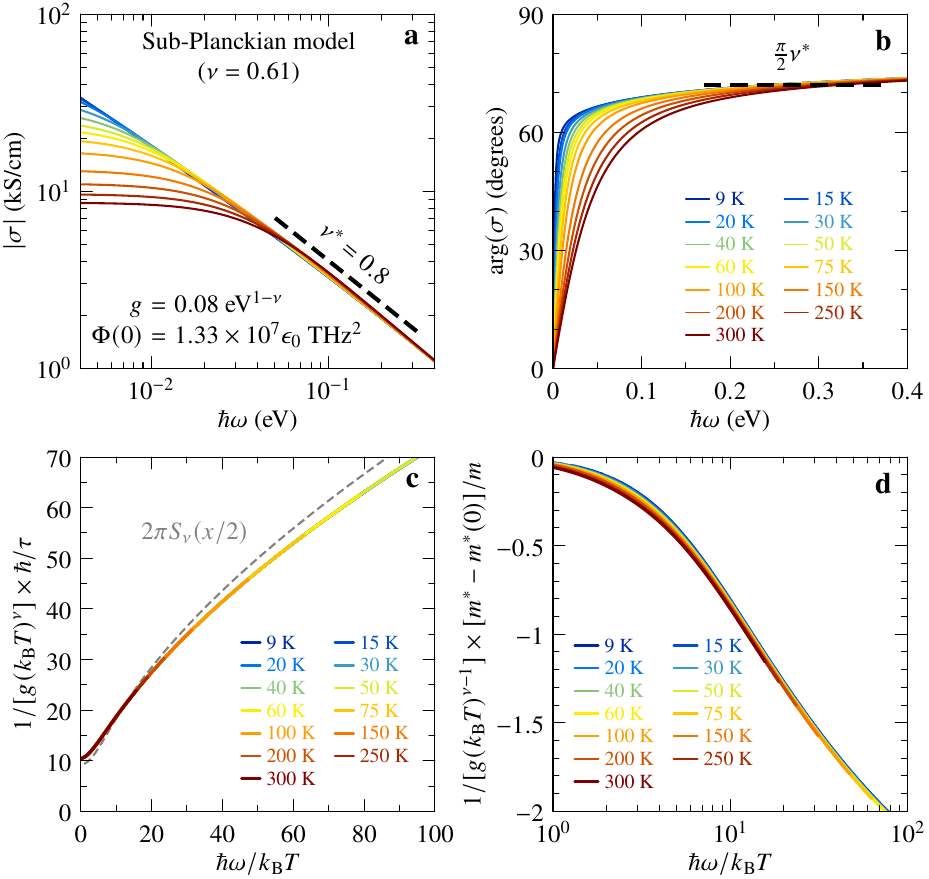}
\caption{\label{fig:fig14}
\textbf{Effective exponent and scaling for the sub-Planckian model.}
\textbf{a} Modulus and \textbf{b} phase of the optical conductivity given by Eqs.~(\ref{eq:sigma}) of the main text and (\ref{eq:Sigma-nu}) with the parameters indicated in \textbf{a}. \textbf{c} Approximate collapse of the scattering rate and \textbf{d} mass enhancement.
}
\end{figure}

In the regimes (II) and (III), we use the approximation Eq.~(\ref{eq:sigma23}), together with the asymptotic self-energy given by Eq.~(\ref{eq:ReSigma-nu}) for the real part and by $-\pi g|\varepsilon|^{\nu}/\Gamma(1+\nu)$ for the imaginary part, yielding
	\begin{equation}\label{eq:sigma23-nu}
		\sigma_{\mathrm{(II, III)}}(\omega)\approx\frac{\Phi(0)}{-i\omega
		+2^{1-\nu}\pi\hbar^{\nu-1} g\frac{1-i\tan(\pi\nu/2)}{\Gamma(1+\nu)}\omega^{\nu}}.
	\end{equation}
This expression reproduces the behavior proportional to $i/\omega$ required by causality in the limit $\omega\to\infty$, irrespective of the value of $\nu$, and interpolates well across $\omega^*$, as shown by the thick lines in \EDF~\ref{fig:fig13}. In the regime (II), a nontrivial power law emerges with an apparent exponent $\nu^*>\nu$. The logarithmic derivative of $|\sigma_{\mathrm{(II, III)}}|$ evaluated at $\omega^*/2$ gives the exponent
	\begin{equation}\label{eq:nustar-nu}
		\nu^*(\nu<1)=\nu+\frac{\left(2^{2\nu-1}+2^{4\nu-2}\right)(1-\nu)}
		{2^{2\nu}+2^{4\nu-2}+1/\sin^2(\pi\nu/2)}.
	\end{equation}
We emphasize that this formula is not valid for $\nu=1$. The cases $\nu=1$ and $\nu<1$ have different analytic structures, with the consequence that the apparent exponent in Eq.~(\ref{eq:nustar-nu}) is independent of the coupling $g$, while in the Planckian case, Eq.~(\ref{eq:nustar}) of the main text, it depends only on $g$. The function (\ref{eq:nustar-nu}) is displayed as an inset in \EDF~\ref{fig:fig13}. An effective exponent of $0.8$, as observed experimentally in LSCO, requires $\nu=0.61$ (see also \EDF~\ref{fig:fig14}). This in turn implies a resistivity $\rho\sim T^{0.61}$, in marked disagreement with experiments, so that a sub-Planckian interpretation is not tenable.

\begin{figure*}[t]
\includegraphics[width=0.8\textwidth]{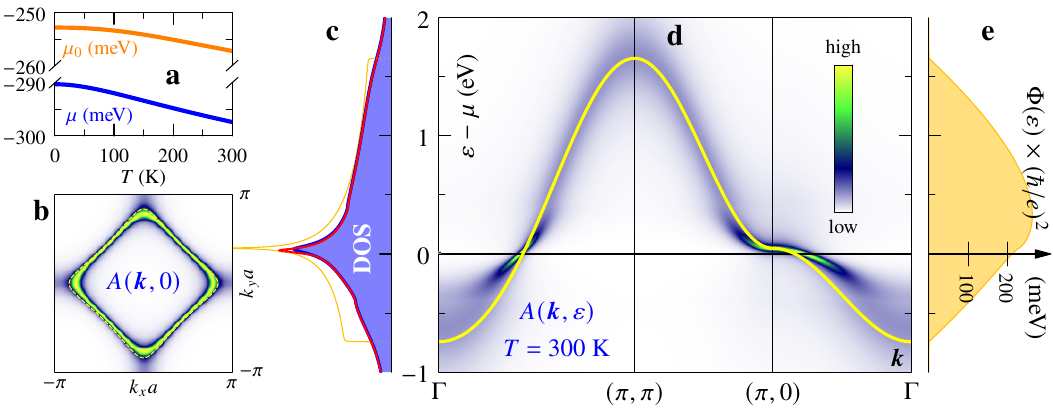}
\caption{\label{fig:fig15}
\textbf{\boldmath Tight-binding model for LSCO at $p=0.24$.}
\textbf{a} Temperature dependence of the noninteracting (orange) and interacting (blue) chemical potential. \textbf{b} Noninteracting Fermi surface (dotted) and zero-energy spectral function at $T=300$~K (same color scale as in \textbf{d}). \textbf{c} Noninteracting (orange) and interacting density of states at $T=10$~K (red) and $T=300$~K (blue). \textbf{d} Tight-binding band (yellow) and spectral function at $T=300$~K. \textbf{e} Transport function. The energies in \textbf{c}, \textbf{d}, and \textbf{e} are measured relative to the corresponding chemical potential.
}
\end{figure*}

\EDF~\ref{fig:fig14}\textbf{a} and \EDF~\ref{fig:fig14}\textbf{b} show the modulus and argument of the optical conductivity calculated numerically using Eqs.~(\ref{eq:sigma}) and (\ref{eq:Sigma-nu}) with $\nu=0.61$. The coupling constant $g$ is fixed such that the magnitude of the conductivity is similar to that in Fig.~\ref{fig:fig3} of the main text. This calculation reproduces the power law observed experimentally in the modulus, but predicts a behavior of the argument that agrees less well with experiment than the model with $\nu=1$, especially at low frequency (compare Figs.~\ref{fig:fig3}\textbf{b}, \ref{fig:fig3}\textbf{d}, and \EDF~\ref{fig:fig14}\textbf{b}). \EDF~\ref{fig:fig14}\textbf{c} and \EDF~\ref{fig:fig14}\textbf{d} show that $1/\tau$ and $m^*$, after being properly scaled (and divided by $g$ to produce dimensionless quantities), display a good collapse as a function of $\omega/T$. This justifies the scaling laws employed in \SIS~\ref{app:scaling} for $\nu<1$. The most striking disagreement with experiment is seen in the scattering rate. Indeed, the scaling requires to normalize $1/\tau$ by $T^{\nu}$, which does not lead to a good collapse of the experimental data for $\nu\approx0.6$ (see \EDF~\ref{fig:fig7}).

\section{Tight-binding model and one-particle properties}
\label{app:tight-binding}

Here, we present the tight-binding model that is used in the main text for fixing the values of the band mass $m$ and spectral weight $K$. We discuss the properties of the model and show how these properties are modified by the self-energy.

The one-particle properties of the model are displayed in \EDF~\ref{fig:fig15}. The tight-binding dispersion with up to second-neighbor hopping amplitudes is given by $\varepsilon_{\vec{k}}=-2t[\cos(k_xa)+\cos(k_ya)]-4t'\cos(k_xa)\cos(k_ya)$ with $t=0.3$~eV, $t'/t=-0.17$, and $a=3.78$~\AA. All properties are shown for an electron density $n=0.76/a^2$, i.e., a hole doping $p=0.24$. The noninteracting chemical potential $\mu_0$ varies from $-253$~meV at $T=0$ to $-257$~meV at $T=300$~K (\EDF~\ref{fig:fig15}\textbf{a}). The corresponding tight-binding band with energies measured relative to $\mu_0(T=0)$ is shown in \EDF~\ref{fig:fig15}\textbf{d}, and the associated density of states in \EDF~\ref{fig:fig15}\textbf{c}. The Fermi-level DOS is $1.646$~eV$^{-1}a^{-2}$, which corresponds to a band mass $m=2.76m_e$. The Fermi surface is closed around the Brillouin-zone center, as seen in \EDF~\ref{fig:fig15}\textbf{b}. The transport function
	\begin{equation}
		\Phi(\varepsilon)=2e^2\int_{\mathrm{BZ}}\frac{d^2k}{(2\pi)^2}
		\left(\frac{1}{\hbar}\frac{d\varepsilon_{\vec{k}}}{dk_x}\right)^2
		\delta(\varepsilon+\mu_0-\varepsilon_{\vec{k}})
	\end{equation}		
is displayed in \EDF~\ref{fig:fig15}\textbf{e}; the $T=0$ value of $(\hbar/e)^2\Phi(0)$ is 211~meV.

The self-energy, Eq.~(\ref{eq:Sigma}) of the main text, renormalizes the dispersion and consequently all one-particle properties. The chemical potential calculated for $g=0.23$ and $\Lambda=0.4$~eV is shown in \EDF~\ref{fig:fig15}\textbf{a} and varies from $-290$~meV at $T=0$ to $-297$~meV at $T=300$~K. We emphasize that the high-energy details of the self-energy influence the value of $\mu$, such that this model calculation, although internally consistent, is not expected to be realistic for LSCO as it misses those high-energy non-universal aspects. The downward renormalization of the chemical potential by the interaction implies a renormalization of the Fermi surface (maximum of the zero-energy spectral function), as seen in \EDF~\ref{fig:fig15}\textbf{b}. A change of the Fermi-surface volume at fixed density seems to violate Luttinger's theorem. The theorem is not applicable here, however, because it requires a true Fermi surface~---~a discontinuity of the momentum distribution. Since the self-energy does not vanish on the Fermi surface, there is no such discontinuity in the model. The color map in \EDF~\ref{fig:fig15}\textbf{d} shows the renormalization of the tight-binding band, which, since the self-energy is local, is the same at all wave-vectors. The corresponding interacting DOS is plotted in \EDF~\ref{fig:fig15}\textbf{c} for two temperatures.

\section{Effect of particle-hole asymmetry}
\label{app:ph-asymmetry}

In the main text, we compare the optical spectra of LSCO with microscopic models that possess particle-hole (p-h) symmetry. The cuprate materials break p-h symmetry at the level of the band structure and, consequently, also at the level of the self-energy. The band-structure effects enter the optical spectra via the energy-dependent transport function $\Phi(\varepsilon)$ replacing $\Phi(0)$ in Eq.~(\ref{eq:sigma}) of the main text. The transport function is smooth (see \EDF~\ref{fig:fig15}\textbf{e}), unlike the DOS that varies rapidly due to the van Hove singularity (\EDF~\ref{fig:fig15}\textbf{c}). As $\Phi(\varepsilon)$ is almost featureless, we do not expect a significant effect of the band structure on the frequency dependence of the optical spectra. These spectra represent transitions from occupied to empty states and thus mix scattering effects at positive and negative energies. For that reason, they should also not be particularly sensitive to the p-h asymmetry of the self-energy. It is indeed clear that the self-energy is to a large extent averaged between positive and negative energies in Eq.~(\ref{eq:sigma}). In order to check this expectation and justify our use of p-h symmetric models for data analysis, we consider here a more general model including p-h asymmetry in the self-energy. This model was proposed recently to investigate the thermopower of non-Fermi liquids \cite{Georges-2021-s}, which is sensitive to p-h asymmetry because, unlike the optical spectra, it vanishes identically for p-h symmetric systems.

\begin{figure}[tb]
\includegraphics[width=0.5\textwidth]{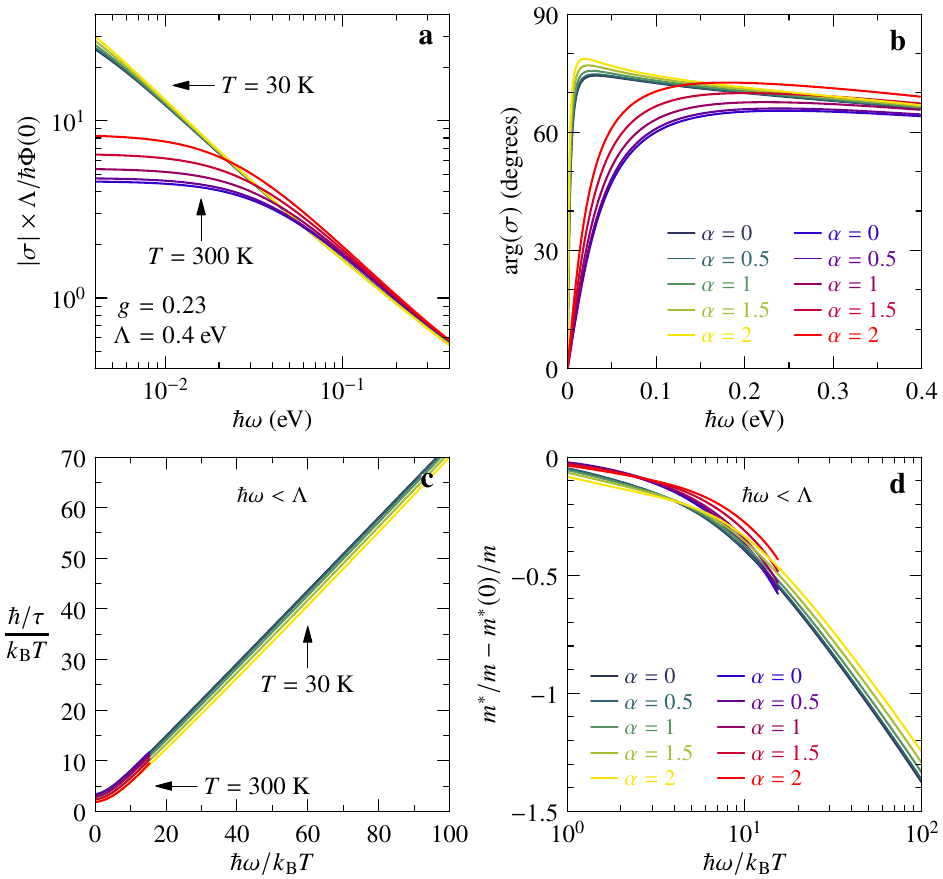}
\caption{\label{fig:fig16}
\textbf{Weak effect of particle-hole asymmetry.}
\textbf{a} Modulus and \textbf{b} phase of the optical conductivity given by Eqs.~(\ref{eq:sigma}) and (\ref{eq:S-nu-alpha}) with $\nu=1$ and the other parameters as indicated in \textbf{a}; the dependence on $\alpha$ is shown for low and room temperature in blue--yellow and blue--red shades, respectively. \textbf{c} Corresponding $\alpha$ dependence of the scattering rate and \textbf{d} the mass enhancement.
}
\end{figure}
Particle-hole asymmetry is introduced in the self-energy by replacing $S_{\nu}(x)$ in Eq.~(\ref{eq:Sigma-nu}) by the function \cite{Georges-2021-s}
	\begin{equation}\label{eq:S-nu-alpha}
		S_{\nu}^{\alpha}(x)=\frac{(2\pi)^{\nu}}{\pi\,\Gamma(1+\nu)}\frac{\cosh(x/2)}{\cosh(\alpha/2)}
		\,\left|\,\Gamma\left(\frac{1+\nu}{2}+i\frac{x+\alpha}{2\pi}\right)\right|^2.
	\end{equation}
A positive value of the dimensionless parameter $\alpha$ skews the scattering by enhancing the scattering rate for holes relative to that for particles. In \EDF~\ref{fig:fig16}, we illustrate the effect of $\alpha$ on the optical conductivity. Results are shown for the Planckian case $\nu=1$. Similar results are found for $\nu<1$. \EDF~\ref{fig:fig16}\textbf{a} and \EDF~\ref{fig:fig16}\textbf{b} show that upon increasing p-h asymmetry from zero to $\alpha=2$, the apparent exponent of the conductivity hardly changes, both at low and room temperatures. Note that $\alpha=2$ represents a strong p-h symmetry violation: with this value, the scattering rate is $\sim7$ times larger for holes at $\varepsilon\ll-\kB T$ than for electrons at $\varepsilon\gg \kB T$. \EDF~\ref{fig:fig16}\textbf{c} and \ref{fig:fig16}\textbf{d} show that the $\omega/T$ scaling laws of the optical scattering rate and mass enhancement also only change in a minor quantitative way upon increasing p-h asymmetry.

\let\oldaddcontentsline\addcontentsline%
\renewcommand{\addcontentsline}[3]{}%

\end{document}